\documentclass[useAMS,usenatbib]{mn2e}
\input{epsf}
\usepackage{amssymb}
\usepackage{natbib}
\usepackage{color}
\usepackage{journals}
\usepackage{multirow}
\usepackage{booktabs,threeparttable}

\newbox\grsign \setbox\grsign=\hbox{$>$} \newdimen\grdimen \grdimen=\ht\grsign
\newbox\simlessbox \newbox\simgreatbox
\setbox\simgreatbox=\hbox{\raise.5ex\hbox{$>$}\llap
     {\lower.5ex\hbox{$\sim$}}}\ht1=\grdimen\dp1=0pt
\setbox\simlessbox=\hbox{\raise.5ex\hbox{$<$}\llap 
     {\lower.5ex\hbox{$\sim$}}}\ht2=\grdimen\dp2=0pt

\definecolor{darkgreen}{rgb}{0.0,0.5,0.0}
\definecolor{darkred}{rgb}{0.5,0.0,0.0}
\definecolor{brown}{rgb}{0.65,.16,0.16}
\definecolor{grey}{rgb}{0.4,0.5,0.6}
\definecolor{darkmagenta}{rgb}{0.55,0.,0.55}
\definecolor{darkorange}{rgb}{1.,0.55,0.}

\newcommand{\hMpc}{{\ifmmode{h^{-1}{\rm Mpc}}\else{$h^{-1}$Mpc }\fi}}
\newcommand{\hGpc}{{\ifmmode{h^{-1}{\rm Gpc}}\else{$h^{-1}$Gpc }\fi}}
\newcommand{\hkpc}{{\ifmmode{h^{-1}{\rm kpc}}\else{$h^{-1}$kpc }\fi}}
\newcommand{\hMsun}{{\ifmmode{h^{-1}{\rm {M_{\odot}}}}\else{$h^{-1}{\rm{M_{\odot}}}$}\fi}}
\newcommand{\Msun}{{\ifmmode{{\rm {M_{\odot}}}}\else{${\rm{M_{\odot}}}$}\fi}}

\title[Galaxy Cluster Mass Reconstruction Project IV]{Galaxy Cluster Mass Reconstruction Project - IV. Understanding the effects 
of imperfect membership on cluster mass estimation}

\author[Wojtak et al.]{R. Wojtak$^{1,2,3}$\thanks{E-mail:
wojtak@dark-cosmology.dk}, L. Old$^{4}$, G. A. Mamon$^{5}$, F. R. Pearce$^{6}$, R. de Carvalho$^{7}$, C. Sif\'on$^{8}$, 
\newauthor M. E. Gray$^{6}$, R. A. Skibba$^{9,10}$, D. Croton$^{11}$, S. Bamford$^{6}$, D. Gifford$^{12}$, 
\newauthor A. von der Linden$^{13}$, J. C. Mu\~{n}oz-Cuartas$^{14}$, V. M\"uller$^{15}$, R. J. Pearson$^{16}$, E. Rozo$^{3, 17}$,
 \newauthor E. Rykoff$^{3}$, A. Saro$^{18}$, T. Sepp$^{19}$ and E. Tempel$^{15,20}$
\\
$^{1}$Dark Cosmology Centre, Niels Bohr Institute, University of Copenhagen, Juliane Maries Vej 30, DK-2100 Copenhagen, Denmark\\
$^{2}$Kavli Institute
for Particle Astrophysics and Cosmology, Stanford University, 452
Lomita Mall, Stanford, CA 94305-4085, USA\\ 
$^{3}$SLAC National Accelerator Laboratory, Menlo Park, CA 94025, USA\\
$^{4}$Department of Astronomy $\&$ Astrophysics, University of Toronto, Toronto, Canada \\
$^{5}$Institut
d'Astrophysique de Paris (UMR 7095: CNRS \& Sorbonne Universit\'e), 98 bis Bd Arago, F-75014 Paris, France\\
$^{6}$School of Physics and Astronomy, University of Nottingham, Nottingham, NG7 2RD, UK\\ 
$^{7}$Instituto Nacional de Pesquisas Espaciais, MCT, S\~{a}o Jos\'e Dos Campos, Brazil\\ 
$^{8}$Department of Astrophysical Sciences, Peyton Hall, Princeton University, Princeton, NJ 08544, USA \\
$^{9}$Science Communication Program, University of California, Santa Cruz, 1156 High Street, Santa Cruz, CA 95064, USA\\
$^{10}$Freelance Science Journalist, San Diego, CA, USA\\
$^{11}$Centre for Astrophysics $\&$ Supercomputing, Swinburne University of Technology,
PO Box 218, Hawthorn, VIC 3122, Australia \\
$^{12}$Department of Astronomy, University of Michigan, 500 Church
St. Ann Arbor, MI, USA\\
$^{13}$Department of Physics and Astronomy, Stony Brook University, Stony Brook, NY 11794, USA\\
$^{14}$Group for Computational Physics and Astrophysics, Instituto de Fisica, Universidad de Antioquia, Medellin, Colombia\\
$^{15}$Leibniz-Institut f\"ur Astrophysik Potsdam (AIP), An der Sternwarte 16, D-14482 Potsdam, Germany\\ 
$^{16}$School
of Physics and Astronomy, University of Birmingham, Birmingham, B15
2TT, UK\\ 
$^{17}$Department of Physics, University of Arizona, Tucson, AZ, 85721\\
$^{18}$INAF-Osservatorio Astronomico di Trieste, via G. B. Tiepolo 11, I-34143 Trieste,  Italy \\
$^{19}$Tarkvara Tehnoloogia Arenduskeskus (STACC), Estonia\\
$^{20}$Tartu Observatory, University of Tartu, Observatooriumi 1, 61602 T\~oravere, Estonia
}

\begin{document}

\maketitle

\begin{abstract}

The primary difficulty in measuring dynamical masses of galaxy clusters from
galaxy data lies in the separation between true cluster members from
interloping galaxies along the line of sight. We study the impact of 
membership contamination and incompleteness 
on cluster mass estimates obtained with 25 commonly used techniques applied
to nearly 1000 mock clusters.
We show that
all methods overestimate or underestimate cluster masses when applied to 
contaminated or incomplete galaxy samples respectively. This appears to be
the main source of the intrinsic scatter in the mass scaling relation. 
Applying corrections based on a prior
knowledge of contamination and incompleteness can reduce the scatter to the
level of shot noise expected for poorly sampled clusters. We establish an
empirical model quantifying the effect of imperfect membership on cluster
mass estimation and discuss its universal and method-dependent
features. We
find that both imperfect membership and the response of the mass estimators
depend on cluster mass, effectively causing a flattening of the estimated -
true mass relation. Imperfect membership thus alters cluster counts determined 
from spectroscopic surveys, hence the cosmological parameters that depend on such counts. 
\end{abstract}


\section{Introduction}

Virtually all studies of clusters of galaxies rely on accurate measurements
of their mass.
Cluster global masses (defined within some physical radius, such as the virial radius,
within which the cluster should lie near dynamical equilibrium), as well as
cluster mass profiles can be extracted by various astronomical techniques
(see below). The primary challenge is that clusters are viewed in projection in the sky: 
our distance estimators are insufficient to enable the reconstruction of a precise 3D mass distribution. 
This projected view of clusters affects all methods of cluster mass
estimation, including the analysis of X-ray observations of the hot diffuse
cluster gas (assumed to be in hydrostatic equilibrium) or of the
statistical shear of galaxy shapes caused by weak
gravitational lensing.
Mass measurements of clusters using (optically-selected) galaxies as kinematic tracers of their
gravitational potential is particularly difficult,  because of the small number of accurate measurements on one hand, and
because of the inevitable confusion, caused by the Hubble flow, with other galaxies (some belonging to
groups or clusters) up to 10 to 20 virial radii along the line of sight
\citep*{Mam2010}. This makes most of the galaxy-based methods for measuring dynamical masses 
particularly prone to the effects of inaccurate membership \citep{Woj2007}. Despite these limitations 
imposed by imperfect membership, many galaxy-based methods of cluster mass estimation have 
been used for a variety of applications. Various case studies based on  kinematics have provided numerous 
constraints on the cluster mass profiles, the mass-concentration relation and the anisotropy of galaxy orbits 
(\citealp{Woj10}; \citealp*{Mun2014}; Mamon et al. 2018, in prep.).

Moreover, clusters are cosmological tools, since the evolution of the cluster mass function depends strongly on
cosmological parameters. In the optical domain, 
these parameters were constrained using scaling relations of mass with
richness for optical galaxy clusters detected in the Sloan Digital 
Sky Survey (SDSS) data \citep{Roz2010} or with velocity dispersion using
clusters detected through the
Sunyaev-Zel'dovich \citep{Sunyaev&Zeldovich70} effect with the South Pole Telescope 
\citep{Bocq2015}. Undoubtedly, the role of galaxy-based methods for constraining cluster mass
profiles on one hand and cosmological parameters on the other 
will grow even more with the advent of large-scale imaging and spectroscopic cosmological surveys such as 
Euclid\footnote{http://sci.esa.int/euclid/} and the Large Synoptic Sky Survey (LSST)\footnote{https://www.lsst.org}. 
Such ``cluster cosmology'' will improve the present constraints on extensions of a standard $\Lambda$CDM model, 
driven thus far primarily by X-ray observations of galaxy clusters, such as modified gravity \citep{Cat2015}, neutrino physics 
\citep{Man2015} or the dark energy equation of state \citep{Man2014}.

Assigning cluster membership to galaxies observed in the cluster field can be
performed in several ways.
First, clusters are selected as galaxy concentrations on the sky or directly
in redshift space.
Moreover, thanks to the prominence of red galaxies in low redshift clusters
combined with the narrowness of the red sequence of galaxies in colour-luminosity
diagrams, one
can select cluster members appearing on a narrow red sequence in
the colour-magnitude diagram \citep[several narrow sequences may signify several
clusters aligned along the line of sight;][]{Glad2000,Ryk2014}. 
Line-of-sight (LOS) velocities provide extra information, and thanks to the Hubble flow,
obvious interlopers can be identified in several methods, such as: (i)  
global $3\,\sigma$ clipping \citep{Yah1977}, (ii)
searching
for gaps in the global LOS velocity distribution \citep{Fad1996}, (iii) 
selecting maximum local absolute LOS velocities, (iv) using best estimates of
the infall velocity
\citep{Har1996}, (v) the escape velocity 
identified again as LOS
velocity gaps now called ``caustics'' \citep{Dia99}, (vi) or a local
$2.7\,\sigma_{\rm LOS}$ clipping \citep*{Mam2013}.\footnote{The factor 2.7
  was optimized by \cite{Mam2010} on a stack of haloes from a cosmological
  simulation.} 
These methods (except caustics) are iterative
(first guessing a virial radius, and, for the latter one,  assumed mass and
velocity anisotropy profiles). But as mentioned
above, all these methods suffer from inevitable contamination from galaxies lying along
the LOS within
10 or 20 virial radii from the cluster, for which the Hubble flow is
insufficient to push them beyond 2.7 or $3\,\sigma$ in the LOS velocity
distribution of the intrinsic cluster members. Stacking 100 haloes from a
cosmological simulation, \cite{Mam2010} found that as many as 23 per cent of objects
lying within the virial cone (or cylinder) lie within the virial sphere
(after filtering out the interlopers with a local $2.7\,\sigma_{\rm LOS}$ criterion). 
A distinctive approach is based on the Friends-of-Friends (FoF) algorithm applied to projected phase space, but this cluster 
finder cannot reach perfect membership: increasing the linking lengths improves completeness at the expense of reliability 
and vice-versa, making it impossible for any combination of linking lengths to jointly achieve over 83 per cent completeness 
and reliability for clusters of estimated mass above $10^{14}\Msun$ (see $Q_{\rm local}$ in Fig. 9 of \citealp{Duarte&Mamon14}). 
Moreover, with optimal linking lengths, over half the FoF clusters of estimated mass $10^{14}\Msun$ turn out to be secondary 
fragments of more massive clusters, and with the most optimal cluster finders as those of \citet{Yang2007} and MAGGIE 
\citep{Duarte&Mamon15} this fraction remains as high as $\sim 15\%$  (Fig. 10 of \citealp{Duarte&Mamon15}).

For most methods, membership assignment cannot be
fully separated from cluster mass measurement. Arguably all algorithms for selecting cluster members involve 
either mass-dependent cuts in projected phase space, e.g. $3\sigma$ clipping, or mass-dependent scales of 
models assigning a probabilistic membership (a galaxy observed at fixed physical distance from the cluster centre 
is more likely a cluster member if its host cluster is more massive than a baseline value). Consequently, 
membership is typically refined in iterative steps based on trial estimates of cluster mass. 
In this sense, both the mass estimate and the membership classification constitute the outputs
of a self-consistent algorithm processing cluster data.

Cluster membership is commonly treated as a binary feature classifying
galaxies into cluster members or interlopers.
An increasingly popular alternative is a probabilistic approach, i.e. where
each galaxy is assigned a membership probability, enabling determination of mass
profile parameters in a self-consistently Bayesian fashion
(e.g. \citealp{Woj09,Woj2013,Mam2013}), as well as the
detection of the gravitational redshift effect 
(\citealp*{Woj2011,Sad2015}; \citealp{Jim2017}) and the spatial
anisotropy of galaxy kinematics \citep{Ski12}. 
 
Despite its growing popularity (e.g. \citealp{Ryk2014}), the probabilistic approach is only
justified when there is an underlying model for the interloping galaxies. For
stacked clusters, quasi-uniform distributions of interloper LOS velocities can be extracted from
cosmological simulations (as proposed by \citealp{Mam2010}). But for individual
clusters, the LOS velocity distribution of interlopers is the sum of many
quasi-Gaussians (each one for a different cluster along the LOS of the
considered one).\footnote{Maximizing the benefits of the probabilistic
  approach for individual clusters can be achieved 
with machine learning algorithms, whose first applications to cluster
mass estimates resulted in noticeable 
improvements \citep{Nta2015,Nta2016}.} 

Despite a rich variety of mass estimators and techniques for selecting
cluster members, the literature lacks extensive studies on how imperfect
membership degrades cluster mass estimates, how scale-dependent properties of galaxy clusters 
\citep[e.g. mass-dependent amount of substructure;][]{Old2018,Car2017,Rob2017} 
affect selection of cluster members and what solutions can mitigate all these effects. 
With the growing role of galaxy-based cluster mass
estimations in future cosmological surveys, it is timely to quantify the
impact of imperfect membership on mass scaling relations in order to envision
the strategies to minimize it. The present work is the first step in addressing these
problems in a systematic way.

We make use of the Galaxy Cluster Mass Reconstruction Project (GCMRP) data
\citep{Old2014,Old2015} to quantify the effect of imperfect membership on
cluster mass measurements. The data include estimates of mass and galaxy
membership for nearly $1000$ mock low-redshift galaxy clusters analyzed by 25 algorithms, 
together with the true masses and galaxy memberships of these clusters. With these data, 
we can quantify, for each method, how mass bias and scatter depend on membership 
incompleteness and contamination.

Our study is a continuation of the GCMRP. A comprehensive framework of the project with 
built-in procedures of data blinding enables us not only to compare individual mass estimation 
methods exploiting a wide range of possible models, scaling relations and numerous algorithms 
for handling various steps of data processing, but also to study generic properties of cluster mass 
measurements in optical observations. In \citet{Old2015}, we quantified performances of all 25 
methods and showed that those based on richness or abundance matching return most precise 
cluster mass estimates, irrespective of intrinsic assumptions of mock galaxy catalogues. 
Exploiting information on dynamical substructure in our follow-up study \citep{Old2018}, we 
demonstrated that all methods systematically overestimate masses of clusters with significant 
substructure. This bias turned out to affect low-mass clusters more strongly than high-mass 
counterparts.

The article is organized as follows. In Section~2, we describe the mock
observations and the cluster mass reconstruction techniques.  In Section~3,
we estimate the level of imperfect membership in terms of contamination and incompleteness,
and quantify its effects on cluster mass estimates returned by each
method,  explaining our adopted methodology of data analysis. We
study the mass dependence of imperfect membership and the response of the cluster
mass estimators in Section~4, quantifying how
imperfect membership modifies the mass scaling relations. We summarize and
conclude in Section~5.

\section{Data and methods}
We base our analysis on tables, generated by the GCMRP
\citep{Old2015}, of estimated mass and galaxy membership for 967
clusters, obtained with 25 different algorithms, together with the true
masses and memberships. 

Two mock galaxy
catalogues were generated for the 25 algorithms: one using
a \emph{semi-analytical model} (SAM), and the other based on a \emph{halo occupation
distribution} (HOD) approach. In this study,
we use the outputs from all 25 mass reconstruction methods, but only for HOD
mock observations, because the true 3D cluster membership 
assigned to galaxies in the SAM catalogue does not conform with the assumed virial 
overdensity of $200\rho_{c}$ defining both cluster masses in the two galaxy catalogues and 
3D cluster membership in the HOD catalogue (see Section 3.1).

The mock catalogues submitted to the algorithms were developed in three
steps: 1) extracting dark matter haloes from a cosmological dark matter simulation; 2) extracting galaxies from the haloes; 3) building a mock galaxy
catalogue.
We describe below each step in more detail (see \citealp{Old2014,Old2015} for
more details).

\subsection{Cosmological simulation}
 The mock observations were generated using the Bolshoi dissipationless
 cosmological simulation based on a flat $\Lambda$CDM cosmological model 
 with the matter density parameter $\Omega_{\rm m}=0.27$, the rms of 
 the density fluctuations $\sigma_{8}=0.82$, the tilt of the primordial power 
 spectrum $n=0.95$ and the dimensionless Hubble constant $h=0.7$. The simulation 
 follows the evolution of $2048^{3}$ dark
 matter particles of mass $1.35\times 10^{8}\,\hMsun$ within a box of side
 length $250\,\hMpc$ \citep{Kly11}. It was run with the ART adaptive
 refinement code with a force resolution of $1\,\hkpc$. The final halo
 catalogues are complete down to circular velocity of $50$~km~s$^{-1}$
 (corresponding to $M_{\rm 200c}\approx1.3 \times 10^{10}\,\hMsun$ with
 $\sim100$ particles per halo).

\begin{table*} 
 \caption{Summary of the cluster mass reconstruction 
 methods}
\begin{center} 
  \tabcolsep 9pt
\begin{tabular}{l l l l l l}
\hline
\multicolumn{1}{c}{Method}&Initial galaxy&Mass estimation&Type of data
&Colour & Reference \\
&  selection   &        &  required     &    info &   \\
(1)     & (2)       & (3)       & (4)       & (5)       & (6) \\
\hline
\textcolor{darkmagenta}{\textbf{PCN}}&Phase space& \textcolor{darkmagenta}{Richness} &Spectroscopy& No & \citet{Pearson2015} \\
\textcolor{darkmagenta}{\textbf{PFN}}*&FoF& \textcolor{darkmagenta}{Richness} &Spectroscopy& No & \citet{Pearson2015}\\
\textcolor{darkmagenta}{\textbf{NUM}} &Phase space& \textcolor{darkmagenta}{Richness} &Spectroscopy& No & Mamon et al. (in prep.) \\
\textcolor{darkmagenta}{\textbf{RM1}}&Red sequence& \textcolor{darkmagenta}{Richness}&Central spectra& Yes & \citet{Ryk2014}\\ 
\textcolor{darkmagenta}{\textbf{RM2}}*&Red sequence& \textcolor{darkmagenta}{Richness} &Central spectra& Yes & \citet{Ryk2014}\\ 
\textcolor{black}{\textbf{ESC}}&Phase space&  \textcolor{black}{Phase space} &Spectroscopy& No &   \citet{Gif2013}\\ 
\textcolor{black}{\textbf{MPO}}&Phase space& \textcolor{black}{Phase space}&Spectroscopy& Yes &   \citet{Mam2013}\\ 
\textcolor{black}{\textbf{MP1}}& Phase space& \textcolor{black}{Phase space}&Spectroscopy& No &   \citet{Mam2013}\\
\textcolor{black}{\textbf{RW}}& Phase space& \textcolor{black}{Phase space} &Spectroscopy& No & \citet{Woj09}\\
\textcolor{black}{\textbf{TAR}}*&FoF& \textcolor{black}{Phase space} &Spectroscopy& No &{\citet{Tem2014}} \\
\textcolor{blue}{\textbf{PCO}}& Phase space& \textcolor{blue}{Radius} &Spectroscopy& No & \citet{Pearson2015}\\ 
\textcolor{blue}{\textbf{PFO}}*&FoF& \textcolor{blue}{Radius} & Spectroscopy& No & \citet{Pearson2015}\\
\textcolor{blue}{\textbf{PCR}}& Phase space&\textcolor{blue}{Radius}&Spectroscopy& No &  \citet{Pearson2015}\\ 
\textcolor{blue}{\textbf{PFR}}*&FoF&\textcolor{blue}{Radius}&Spectroscopy& No &   \citet{Pearson2015}\\
\textcolor{darkgreen}{\textbf{MVM}}*&FoF& \textcolor{darkgreen}{Abundance matching} &Spectroscopy& No & \citet{Mun2012}\\
\textcolor{darkorange}{\textbf{AS1}}&Red sequence&\textcolor{darkorange}{Velocity dispersion} &Spectroscopy& No & \citet{Sar2013}\\
\textcolor{darkorange}{\textbf{AS2}}&Red sequence&\textcolor{darkorange}{Velocity dispersion}&Spectroscopy& No & \citet{Sar2013}\\
\textcolor{darkorange}{\textbf{AvL}}& Phase space&\textcolor{darkorange}{Velocity dispersion}&Spectroscopy& No & \citet{Linden2007}\\ 
\textcolor{darkorange}{\textbf{CLE}}& Phase space&\textcolor{darkorange}{Velocity dispersion}&Spectroscopy& No & \citet{Mam2013}\\ 
\textcolor{darkorange}{\textbf{CLN}}& Phase space&\textcolor{darkorange}{Velocity dispersion}&Spectroscopy& No & \citet{Mam2013}\\
\textcolor{darkorange}{\textbf{SG1}}& Phase space&\textcolor{darkorange}{Velocity dispersion}&Spectroscopy& No &\citet{Sifon2013}\\
\textcolor{darkorange}{\textbf{SG2}}& Phase space&\textcolor{darkorange}{Velocity dispersion}&Spectroscopy& No & \citet{Sifon2013}\\
\textcolor{darkorange}{\textbf{SG3}}& Phase space&\textcolor{darkorange}{Velocity dispersion}&Spectroscopy& No & \citet{Lopes2009}\\
\textcolor{darkorange}{\textbf{PCS}}& Phase space&\textcolor{darkorange}{Velocity dispersion}&Spectroscopy& No & \citet{Pearson2015} \\ 
\textcolor{darkorange}{\textbf{PFS}}*&FoF&\textcolor{darkorange}{Velocity dispersion}&Spectroscopy& No &\citet{Pearson2015} \\ 
\hline 
\end{tabular}
\end{center}
\parbox{16cm}{Notes: Columns are
  (1): Acronym of algorithm;
  (2): Physical property for initial selection of cluster members;
  (3): Physical property for mass estimation from the initial membership;
  (4): Observed properties;
  (5): Use of colour information;
  (6): Reference. The colours indicate five main classes of cluster mass estimation methods. 
  Acronyms denoted with an asterisk
 indicate that the method did not use our initial object
 target list but rather performed an independent cluster search and matched
 the cluster  
 locations at the end of their analysis.
 See Table~\ref{tab:appendix_table_1} in
 the appendix for more details on each method.}
\label{tab:basic_methods}
\end{table*}

 Dark matter haloes were found using the
 ROCKSTAR algorithm \citep{Beh2013}. The
 halo finder operates in full 6D phase space, enabling it to resolve more
 effectively haloes with spatially aligned centres.  It has been shown to
 recover halo properties with high accuracy and returns halo catalogues that
 are broadly consistent with other halo finders \citep{Kne2011}. Halo masses 
 were calculated using
 a spherical overdensity threshold fixed at $200$ times that of the critical
 density at the considered redshift. These overdensities were estimated
 considering all particles and substructures contained in the halo.
 
\subsection{3D galaxy catalogues}
A three-dimensional galaxy catalogue was generated using an updated HOD model
described in \citet{Ski2006} and \citet{Ski2009}. In this approach, dark
matter haloes are populated with galaxies whose luminosities and colours are
assigned so that the simulated galaxy population approximately
reproduces the observed luminosity function, colour-magnitude distribution,
luminosity- and colour-dependent two-point correlation function measured from
the SDSS.
All galaxy properties are also assumed to be
fully determined by the parent halo mass. During 
the course of the GCMRP, several improvements
regarding phase-space distribution of satellite galaxies were developed and
implemented. All modifications account for a number of effects which are
present in realistic groups or clusters of galaxies, but were neglected in the first 
phase of the project, such as: non-central
positions of brightest cluster galaxies, central galaxy velocity bias
\citep{Ski2011a} or difference between the concentrations of dark matter vs. 
satellite galaxies \citep{Woj2013} as well as those of red vs. blue galaxies 
\citep{Coll2005,Cav2017}.  However, the HOD has two important simplifications:
1) halos are truncated at the virial radius, whereas the 1-halo term of
galaxy clusters extends to beyond 10 virial radii \citep{Trevisan+17}, and 2)
galaxies within the halos are not in local dynamical equilibrium (some of the
algorithms assume this equilibrium).  For a complete description of all
implemented improvements we refer the reader to \citet{Old2015}.

\subsection{Mock galaxy catalogue}
The light cone was produced using online tools of the Theoretical Astrophysical Observatory 
\citep[TAO,][]{Ber2016}. It subtends 60$^{\circ}$ by 60$^{\circ}$ on the sky and covers redshift 
range of $0<z<0.15$. The galaxy sample included in the cone is complete down to a minimum $r$-band 
luminosity of $M_{r}=-19+5\log h$ in the input galaxy catalogue.
The input provided to the algorithms consisted of the full galaxy catalogue
(sky position and redshift), as well as the cluster centres (sky positions
and redshifts) given by the locations of the brightest cluster galaxies. In other words, the GCMRP 
assumes that clusters are previously detected and their centres are known.

\subsection{Mass reconstruction methods}

The 25 algorithms of cluster mass reconstruction exhaust nearly all possible
ways of inferring cluster masses from galaxy
data. Table~\ref{tab:basic_methods} presents a brief
overview of all basic characteristics. Following \citet{Old2015}, we divide the methods into five broad
categories with respect to what data features are effectively used by the
mass estimators: methods based on
cluster richness (richness), radial scale of galaxy overdensity (radius), 
velocity dispersion of galaxies (velocity dispersion),
their distribution in projected phase-space (phase space),
and
abundance matching (equating the cumulative galaxy luminosity function with
the cumulative theoretical
halo mass function).
The main motivation of introducing these categories is to check
for any possible subtrends in the analyzed effects.

The algorithms for selecting cluster members form a part of the whole data
processing. In most cases, they are combined with the mass estimators through
various iterative procedures aimed at refining both mass estimates and
cluster membership. Every method begins with an initial selection of
galaxies, performed in several possible ways:
colour-magnitude space (targeting red sequence
galaxies), projected phase space (cuts in cluster-centric distances and/or
velocities) or by applying a Friends-of-Friends grouping algorithm. 
Table~\ref{tab:basic_methods}  summarizes the initial galaxy selection 
adopted by each method. More detailed information regarding this matter 
is provided in Table~\ref{tab:appendix_table_1} and in the main articles of the 
GCMRP: \citealp{Old2014,Old2015}.

Since no method can recover the exact masses of clusters, one can
think of each method's output
as a recovered vs. true cluster mass scaling relation or equivalently a scaling relation
between mass bias and true mass.

\section{Imperfect Membership}
\subsection{Definitions and raw results}
We adopt a simple and intuitive definition of the cluster membership and
assume that all galaxies within the virial sphere are regarded as cluster
members. We consider the virial radius that of a sphere enclosing a density 
which is 200 times the critical density of the Universe at the redshift of the cluster. 
We quantify the observational selection of cluster members in terms of
contamination $C$ and incompleteness  
$I$. The former measures the relative fraction of interlopers in the selected
galaxy sample, whereas the latter  
measures the fraction of true cluster members that are missing in the sample
of selected galaxies, i.e. 
\begin{equation}
C=\frac{N_{\rm sel,non-mem}}{N_{\rm sel}},
\end{equation}
\begin{equation}
I=\frac{N_{\rm non-sel,mem}}{N_{\rm mem}},
\end{equation}
where $N_{\rm mem}$ is the number of true cluster members, $N_{\rm sel}$ is
the number of  
selected galaxies, $N_{\rm sel,non-mem}$ is the number of selected
interlopers (non-members) and  
$N_{\rm non-sel,mem}$ is the number of missing cluster members. Contamination
and incompleteness  
take values between 0 and 1. For a perfect membership assignment with no
interlopers and all true  
cluster members included, $C=0$ and $I=0$.

\begin{figure*}
\begin{center}
    \leavevmode
        \epsfxsize=15.8cm
    \epsfbox[55 70 800 1055]{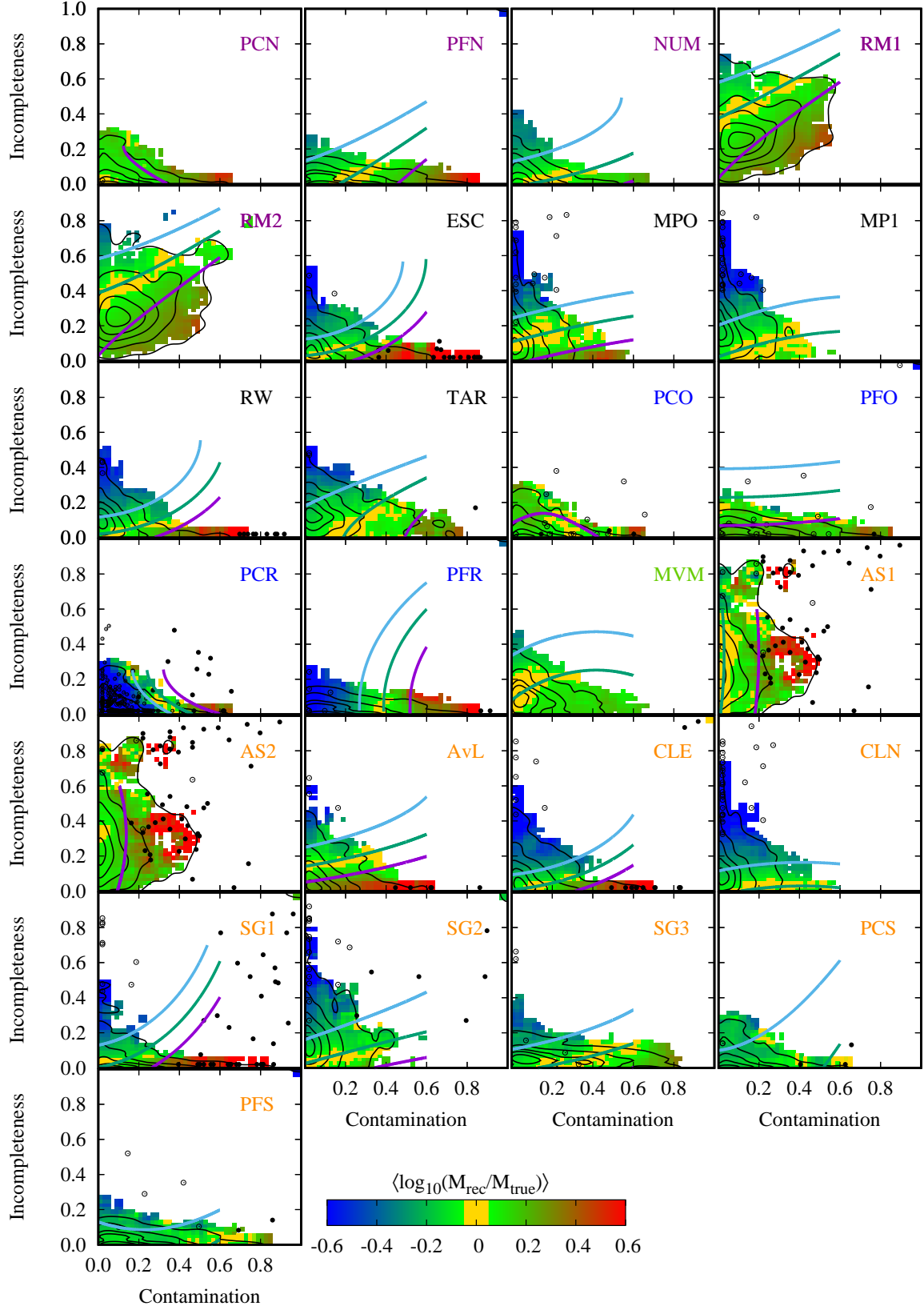}
\end{center}
\caption{Mass bias and frequency as a function of contamination and incompleteness of
  galaxy samples selected by 25 
  different methods for nearly $1000$ mock galaxy clusters. The \emph{black
    contours} show the distribution of contamination and incompleteness
  (isodensity contours containing 10, 30, 50, 70 and 90 per cent of
  clusters). The \emph{colour maps} show the mean mass bias, $\mu = \left\langle\log_{10}\left(M_{\rm rec}/M_{\rm
    true}\right)\right\rangle$, computed using a moving average method with a
  square-shaped $0.1\times0.1$ window function (yellow indicates $|\mu|<0.05$). The \emph{filled} and
  \emph{empty black circles} show catastrophic cases with the estimated
  cluster masses respectively larger or smaller than the corresponding true
  cluster masses by a factor of 10. Pixels containing less than $5$ galaxy clusters within the window were left blank.
  The \emph{coloured lines} show the
  best-fit mass bias model 
  (eq.~[\ref{mu_model}]) with \emph{purple}, \emph{green} and \emph{cyan}
  respectively corresponding to $\mu = 0.2$, 0, and --0.2 (all three present
  unless the range of 
  $\mu$ does not include any of the three fixed values).
  }
\label{CI-HOD2}
\end{figure*}

The black contours in Fig.~\ref{CI-HOD2} show the distributions of
contamination and incompleteness levels in galaxy samples selected by 
the 25 algorithms. The extent of the contours in the figure demonstrates 
that a typical galaxy sample returned by
virtually every method is contaminated and incomplete to some degree. Perfect
membership defined within the 3D virial sphere is in practice attainable only for
a very small fraction of clusters which are most likely isolated. 
It is also apparent that contamination and incompleteness
vary substantially between individual clusters. As we shall demonstrate
below, this has a strong impact both on the accuracy and precision of
cluster mass measurements.

Although Fig.~\ref{CI-HOD2} demonstrates quite substantial differences
between all 25 methods, it is possible to draw some general
conclusions. First, it is clear that cluster membership assignment is more precise when
galaxy velocities are considered (e.g. the large extent of the contours of
the photometric RM1 and RM2 methods compared to the other ones). 
Secondly, despite quite significant differences in how
algorithms handle the selection of cluster members, many methods return
strikingly similar galaxy samples. For example, methods NUM, AvL, ESC, and RW
show fairly similar contours of membership quality, despite their differences
in 
galaxy selection: $2.7\,\sigma_{v}$ for NUM, $2\,\sigma_v$ for AvL, and
escape velocity for ESC and RW. Despite the fact that the former two methods shrink
velocity envelopes by $30$ to $50$ per cent with respect to escape velocity,
they do not return galaxy samples with noticeably smaller contamination. In
fact, all methods, regardless of the employed amplitude of the maximum
velocity 
profile, are similarly affected by a number of clusters with highly
contaminated galaxy samples (see the long horizontal branches at high
contamination and small incompleteness). This demonstrates that attempting
to reduce contamination by narrowing down the initial velocity range does not
guarantee an improvement in cluster membership and should
thus be treated with caution.

\subsection{Effects of imperfect membership on cluster mass estimation}

All methods of cluster mass estimation rely on the assumption of
self-similarity in cluster data. The different mass estimation algorithms
differ in the data feature that is utilized (richness, radius, velocity dispersion,
projected phase-space distribution, abundance matching etc.) and how it scales with cluster mass
(empirical relations or fundamental principles such as the virial
theorem or the Jeans equation of local dynamical equilibrium).
Some methods employ more sophisticated models which effectively
provide higher order corrections to the underlying recovered vs. true cluster mass scaling relations.
Marginalization over nuisance parameters of these models (e.g.,
the shape of the mass density profile for methods based on richness, the velocity
anisotropy in dynamical models) is expected to provide more 
accurate mass determinations.

The apparent scatter in contamination and incompleteness of galaxy samples
selected by each method, highlighted in Fig.~\ref{CI-HOD2}, breaks the
assumption of self-similarity in the input cluster data. This should affect
cluster mass estimation in two ways.
First, this will increase the scatter in the recovered vs. true cluster mass scaling relation, for example if recovered mass depends on richness. 
Secondly, if either galaxy selection or the response of the mass estimator depends on true cluster mass, 
then one should also expect an alteration of the slope of the recovered vs. true cluster mass scaling relation.

To begin with, we neglect a possible mass-dependence of membership assignments and mass estimators, 
and focus on the global relationship between cluster mass accuracy and imperfect membership. 
The main effects of imperfect membership can then be assessed
by comparing the mean difference between recovered 
(estimated) and true log masses, $\log_{10}(M_{\rm rec}/M_{\rm true})$, 
as a function of contamination and incompleteness. The results are shown in Fig.~\ref{CI-HOD2} as colour maps.

The colour gradient apparent in nearly all panels of Fig.~\ref{CI-HOD2} 
demonstrates that virtually all methods tend to 
overestimate (underestimate) cluster masses with higher contamination
(incompleteness) of selected galaxy samples. In most cases, the maps display
distinct lines of degeneracy along which an overestimation due to
contamination is compensated by an underestimation caused by
incompleteness. This feature is quite intuitive for richness-based methods
where the mass proxy is simply proportional to the number of cluster
members. For methods utilizing information on velocities, similar trends
appear to be naturally expected too if we realize that galaxy selection
operates effectively in the tails of the velocity distribution of cluster
members: any contamination or incompleteness of galaxy samples in this
velocity regime automatically leads to an overestimate or underestimate of
the velocity dispersion and consequently the cluster mass.

Comparing the black contours and the colour maps in Fig.~\ref{CI-HOD2}, we
can see that in most cases the maximum of the accuracy (yellow) does not coincide 
with the peak in the membership distribution. This indicates that cluster mass estimates 
may be biased even though galaxy samples are characterized by contamination 
and incompleteness typical of a given method. The mass bias corresponding to the peak of 
the distribution (the innermost isocontour in the figure) varies between the methods 
with median and scatter of $-0.05$~dex and 0.16~dex.

\citet{Old2015} showed that most methods return a fraction (with mean of 3 per cent) 
of mass estimates which deviate from the true masses by as much as a factor of 10. 
Highlighting these catastrophic cases as circles in Fig.~\ref{CI-HOD2}, we identify them
with incidents of extreme contamination or incompleteness.

\subsection{Impact on cluster mass estimation: the model}
Fig.~\ref{CI-HOD2} demonstrates that the mean logarithmic differences between the recovered and 
true masses depends on contamination and incompleteness, i.e.
\begin{equation}
\langle \log_{10}(M_{\rm rec}/M_{\rm true})\rangle =\mu (C,I).
\end{equation}
The function $\mu(C,I)$ encapsulates our base model describing the primary effect of imperfect membership 
on cluster mass estimation. In the following, we make use of the output of each method to determine 
its empirical approximations.

Assuming a Gaussian distribution of variable $x=\log_{10}(M_{\rm rec}/M_{\rm
  true})$, the $\mu(C,I)$ of a given algorithm 
can be found by maximizing the following likelihood function:
\begin{equation}
L\propto \prod_{i} [(1-w_{\rm c})G(x_{i}; \mu(C,I), \sigma)+w_{\rm c}G(x_{i};
  \mu(C,I), \sigma_{\rm c})] \ ,
\label{like1}
\end{equation}
where $G(x; \mu, \sigma)$ is a Gaussian function of $x$ with mean $\mu$ and
variance $\sigma^{2}$ and where the product is over the 967 clusters analyzed by
the algorithm. The second 
term in the likelihood accounts for outliers whose relative fraction in the cluster sample is described by 
nuisance parameter $w_{\rm c}$. We assume a flat distribution of outliers by fixing $\sigma_{\rm c}$ 
to a large value.

We optimize the parametric form of $\mu(C,I)$ by considering a series of truncated Taylor expansions 
about the mean contamination $\langle C\rangle$ and mean incompleteness $\langle I\rangle$. Employing the 
Bayesian Information Criterion (BIC) for model selection, we found that the following parametrization
\begin{eqnarray}
\mu=\mu_{0} & + & \mu_{C1}(C-\langle C\rangle)+\mu_{C2}(C-\langle C\rangle)^{2} \nonumber \\
    & + & \mu_{I1}(I-\langle I\rangle)+\mu_{I2}(I-\langle I\rangle)^{2}
 \label{mu_model}
\end{eqnarray}
is sufficient to provide a satisfactory description of all data sets. The
majority of the methods, including all higher ranking methods
\citep[see][]{Old2015}, do not support more complex models with cross or
higher order terms (15 methods favor model [\ref{mu_model}] over a purely linear model with 
$\Delta\rm{BIC}<-6$, while only 2 methods favor the inclusion of the second order cross 
term).

\begin{figure}
\begin{center}
    \leavevmode
    \epsfxsize=8.5cm
    \epsfbox[70 50 480 430]{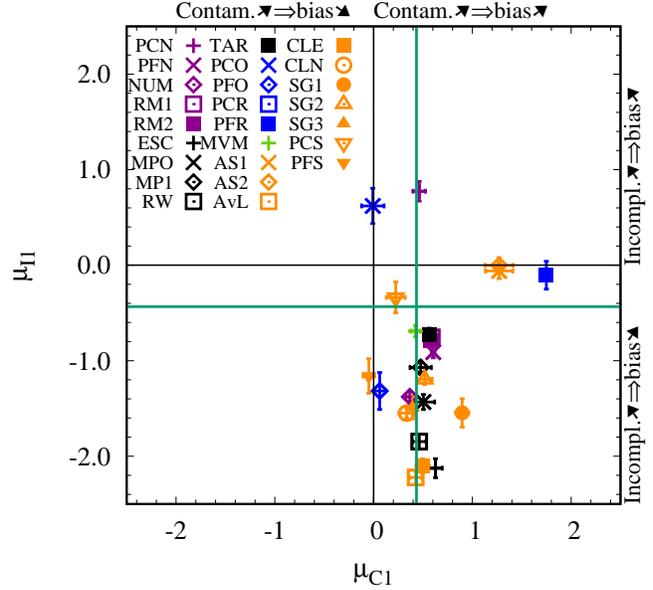}
\end{center}
\caption{Increase of cluster mass biases of the 25 algorithms on the contamination and
  incompleteness of the galaxies selected by them in their mass
  analysis. The horizontal (vertical) axis shows the linear slope of the 
  relationship between bias $\langle \log_{10}(M_{\rm rec}/M_{\rm true})\rangle$ and contamination 
  (incompleteness) about its mean value, i.e. 
  $\mu_{C1}=\delta\langle \log_{10}(M_{\rm rec}/M_{\rm true})\rangle/\delta C$ 
  and $\mu_{I1}=\delta\langle \log_{10}(M_{\rm rec}/M_{\rm true})\rangle/\delta I$. 
  These slopes are determined by fitting the model given by eq.~(\ref{mu_model}). 
  The \emph{green vertical} and \emph{horizontal lines} indicate the coefficients expected 
for an idealized richness-based estimator employing the number of cluster members as a proxy 
for cluster mass. Method PCR lies out of the bounds of the plot with $\mu_{C1}\approx\mu_{I1}\approx3$.
\label{lin_slopes}
}
\end{figure}

We determine best fit parameters and the corresponding confidence ranges
using a Monte Carlo Markov Chain (MCMC) technique based on the
Hastings-Metropolis algorithm. We assume that the effective variance in the
first Gaussian term of the likelihood (eq.~[\ref{like1}]) consists of the
intrinsic scatter and the contribution from shot (Poisson) noise, e.g.
\begin{equation}
 \sigma^2 = \sigma_0^2 + \left( { 100\over N_{\rm true} } \right)\,\sigma_1^2
\label{scatter}
\end{equation}
where $N_{\rm true}$ is the number of true cluster members, and where the
irreducible scatter, $\sigma_0$, and the richness-dependence of the scatter,
$\sigma_1$, are both treated as additional free parameters in the MCMC analysis. 
As we shall see in Section~\ref{sec:scatterM}, $\sigma_0$ and
$\sigma_1$ are both strongly affected by imperfect membership. 

For each method, it is possible to find a combination of contamination and incompleteness for which the mass overestimation 
due to contamination is fully compensated by its counterpart due to incompleteness. 
The cyan, green and purple curved lines in Fig.~\ref{CI-HOD2} show the best
fitting models found for each method. The models 
are presented in the form of lines of constant $\mu$ with
$\mu=0,\pm0.2$. Goodness of fit is addressed in more detail in Appendix~\ref{sec:goodness},
where we show residuals for each method (see Fig.~\ref{CI-HOD2_res}).

The linear terms of the model provide an accurate approximation in a narrow range of 
contamination and incompleteness about their mean values. 
Reducing the comparison between the methods to the level of linear coefficients $\mu_{C1}$
and $\mu_{I1}$, as shown in Fig.~\ref{lin_slopes}. The figure shows that most mass estimators respond to changes in
contamination in nearly the same way.  Neglecting 7 outliers lying outside of 
the $\pm3\sigma$ range (AS1, AS2,
PFR, PFS, PCO, PFO, PCR), we find that the coefficients for the remaining
methods can be described by a remarkably narrow distribution with mean of
0.50 and scatter 0.12.  This distribution of $\mu_{C1}$ is consistent with the expectation
of a simple, idealized richness-based mass estimator, where the recovered
mass varies as the number of
estimated cluster members, i.e.
$M_{\rm rec}\sim N_{\rm sel, mem}$ and $M_{\rm true}\sim N_{\rm mem}$,
leading to
$\mu =  \log_{10}(1+C) \simeq C/(\ln 10) \simeq 0.43\,C$ (see the vertical
green line in Fig.~\ref{lin_slopes}). This simple model appears to universally describe the effect of
contamination on the mass estimation in three distinct groups of methods
based on richness, velocity dispersion, and distribution in projected phase-space.
The 7 outliers are among the methods with the lowest merit of the
mass recovery accuracy \citep[see][]{Old2015}.

The analogous simple, idealized, model for incompleteness, would lead to
$\mu =  \log_{10}(1-I) \simeq -I/(\ln 10) \simeq -0.43\,I$.
However, the nearly universal response of the mass estimators to changes in
contamination does not have its analogy for incompleteness, as 
coefficient $\mu_{I1}$ ranges from $-2.3$ to $0$ (neglecting three outliers
with $\mu_{I1}>0$). Most methods are characterized by coefficients $\mu_{I1} < -0.43$ 
(see the green horizontal line in Fig.~\ref{lin_slopes}). The lowest
$\mu_{I1}$, coefficients indicating the strongest dependance of the mass estimators on
incompleteness, are found for methods based on velocity dispersion and on
the distribution in projected phase-space.
This sensitivity of kinematical methods
to incompleteness is not surprising: these methods select galaxies in 
velocity space; therefore, the missing cluster members are most likely to lie
in the tails of the velocity distribution, in contrast to interlopers whose
velocity distribution resemble quite closely the velocity distribution of
true cluster members \citep[][]{Mam2010}. This in turn leads to a stronger
effect on the velocity dispersion and the corresponding cluster mass estimate
due to incompleteness compared to that due to contamination.

\section{Mass-dependent effects of imperfect membership on mass estimation}

\begin{figure}
\begin{center}
    \leavevmode
    \epsfxsize=8.5cm
    \epsfbox[55 55 500 435]{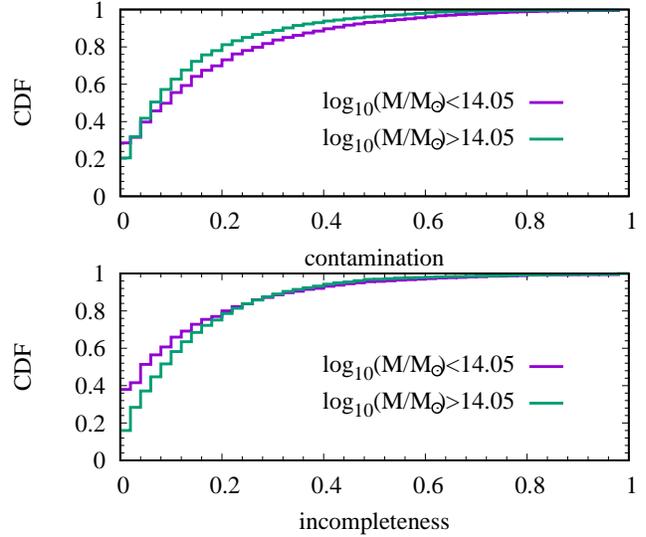}
\end{center}
\caption{Cumulative distribution functions of contamination (\emph{top
    panel}) and incompleteness (\emph{bottom panel}) in the high-mass (\emph{cyan}) and
  low-mass (\emph{purple}) clusters (splitting the true cluster masses by the median value of 
$\log_{10}(M_{\rm true}/{\rm M}_\odot)=14.05$).
The probability combines results from all 25 methods and thus represents a
global trend.
Selected galaxy 
samples tend to be more incomplete for the most massive clusters and more contaminated for the least massive 
clusters. 
}
\label{cpdf}
\end{figure}

Imperfect membership may give rise to a mass-dependent effect on cluster mass
estimation in two different ways.  First, the algorithms selecting cluster
members may be scale-dependent and return galaxy samples with contamination
and incompleteness that depend on cluster mass. Second, there is no guarantee
that the same level of contamination or incompleteness affects cluster mass
measurements in the same way regardless of the actual cluster mass.  This
kind of mass dependance may occur due to the presence of an exponential
cut-off in the mass function that breaks the self-similarity between how small
galaxy groups perturb mass measurements of massive galaxy clusters and vice
versa.  In the following subsections, we seek to identify the extent to which these 
two aspects of mass-dependent imperfect membership underlie the performance 
of all methods.

\subsection{Mass-dependent galaxy selection}

To compare the effects of cluster mass on the relation between mass bias and
membership quality, we compare contamination and incompleteness levels in two subsamples of
respectively low and high true cluster mass, splitting at the median mass of
$\log_{10}(M_{\rm true}/{\rm M}_\odot)=14.05$.  Fig.~\ref{cpdf} shows the
cumulative probability of contamination and incompleteness in the two groups
of clusters. The calculation combines the data from all methods and thus the
results demonstrate a global trend common to all 25 techniques ($25\times500$ 
data points used for calculating each cumulative distribution). The
distribution of contamination (incompleteness) in the two samples of galaxy
clusters appears to be significantly different. The maximum difference between 
the cumulative probability distributions is $0.09$ for contamination and $0.16$ 
for incompleteness, while the upper limit required for rejection of the null hypothesis 
at level $p=0.001$ of the Kolmogorov-Smirnov test is $0.025$. Galaxy samples selected 
from high-mass clusters tend to be less contaminated, but more 
incomplete.

\begin{figure}
\begin{center}
    \leavevmode
    \epsfxsize=8.5cm
    \epsfbox[70 50 480 430]{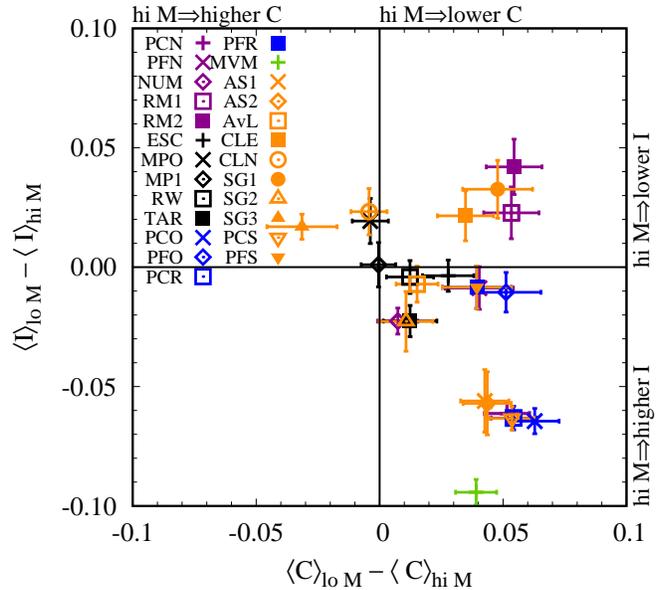}
\end{center}
\caption{Differences in low vs. high-mass (lo M, hi M) cluster contamination (C) and
  incompleteness (I), where the cluster masses are split by the median true mass of
  $\log_{10}(M_{\rm true}/{\rm M}_\odot)=14.05$.
  Most methods return more contaminated samples for less
  massive clusters (21 methods) and more incomplete samples for more massive
  ones (17 methods).
}
\label{CI-differences}
\end{figure}

The global trend shown in Fig.~\ref{cpdf} reflects in large part the behavior of
each method analyzed separately.  This is demonstrated in
Fig.~\ref{CI-differences} which shows the differences between the two groups
of clusters in terms of the mean contamination and incompleteness calculated
for each method (with errors estimated from bootstrapping). Except for four
methods (MPO, MP1, CLN and SG3), the mean contamination for the least massive
clusters is clearly larger than that for the most massive ones at 
confidence levels ranging from $1\,\sigma$ up to $5\,\sigma$. 
In agreement with Fig.~\ref{cpdf}, one also notices that galaxy samples selected from more 
massive clusters appear to be more incomplete, although 
the number of exceptions increases here to 8 out of 25 methods. Among 
all five groups of the methods, those based on projected-phase-space analysis appear to 
minimize the dependance of galaxy selection on cluster mass.

The mass dependence of cluster membership can also be demonstrated by using the galaxy sample 
selection function which is defined as the ratio of completeness to purity, where completeness~$=1-$~incompleteness 
and purity~$=1-$~contamination \citep[by analogy to the cluster selection function; see e.g.][]{Soa2011}. 
Fig.~\ref{CP-HOD2} in the appendix shows the selection functions for all 25 methods compared to 
mean biases of cluster mass estimates. The strongest 
sensitivity of the mass function to cluster mass is unsurprisingly revealed by the methods with reversed 
dependences of contamination and incompleteness on cluster mass (bottom right corner in 
Fig.~\ref{CI-differences}), such as MVM, PCO, PCR, PCS, AS1 and AS2. As expected from the 
trends shown in Fig.~\ref{CI-differences}, these methods are characterized by the selection functions that 
decrease with increasing cluster mass. Fig.~\ref{CP-HOD2} also points to another aspect of interconnections 
between membership and cluster mass measurements. The contours of constant bias appear not to coincide 
precisely with the contours of constant selection function (see e.g. NUM, PCO and MVM as extreme examples). 
This implies that some methods may return cluster mass estimates with mass-dependent bias even though 
selection of cluster members remains strictly independent of cluster mass. We quantify this effect in a more rigorous 
way in the following section.

\subsection{Mass-dependent effects on mass estimation}
We quantify whether the impact of imperfect membership on the mass bias depends on the cluster mass 
by analyzing the output data of all methods using the following
generalization of our base model (\ref{mu_model}): 
\begin{eqnarray}
  \mu&\!\!\!\!=\!\!\!\!&\mu_{0}  +  (\alpha_{0}-1)\log_{10}(M_{\rm
    true}/M_{0}) \nonumber \\ 
  &\mbox{}& \quad + (M_{\rm true}/M_{0})^{\alpha_{\rm
      C}}\,[\mu_{C1}\,(C\!-\!\langle C\rangle)+\mu_{C2}\,(C\!-\!\langle
    C\rangle)^{2}] \nonumber \\ 
  &\mbox{}& \quad +  (M_{\rm true}/M_{0})^{\alpha_{\rm
      I}}\,[\mu_{I1}\,(I\!-\!\langle I\rangle)+\mu_{I2}\,(I\!-\!\langle
    I\rangle)^{2}]\ . 
 \label{mu_model_g}
\end{eqnarray}
The three new mass-dependent terms account for a potential
dependance of three different components of the base model on cluster
mass. The first new term (proportional to $\alpha_{0}-1$) describes an
alteration of the slope of the scaling relation between the recovered and
true cluster mass. When ignoring all terms dependent on contamination and
incompleteness, the $M_{\rm rec}\propto M_{\rm true}^{\alpha_{0}}$ scaling relation 
absorbs all effects of imperfect membership. As measured by \citet{Old2015}, 
the slope $\alpha_{0}$ in this case varies quite substantially between the methods 
with mean and scatter of $0.97$ and $0.19$. In our approach, $\alpha_{0}$ is determined
simultaneously with all parameters of the model accounting for 
imperfect membership. Comparing its values to those from \citet{Old2015} will
demonstrate if and how contamination and incompleteness modifies the $M_{\rm
  rec}-M_{\rm true}$ scaling relation. The second and third new terms
describe a mass-dependent response of mass estimators to contamination and
incompleteness. We choose a power-law ansatz in order to avoid a sign change
for the expressions in square brackets and minimize degeneracies with parameters of
the base model. The signs of the slopes distinguish between whether the
effect of contamination or incompleteness is stronger in more massive
clusters (positive signs) or less massive ones (negative signs). The pivot
mass $M_{0}$ in all three new terms is set at the median mass of the whole
cluster sample, i.e. $\log_{10}M_{\rm true}=14.05$. We find best fit
parameters following the same approach as outlined in Section 3.2. 
Table~\ref{tab:best_fit} in the appendix shows 
the results for each method and compares to those obtained for a simplistic model 
neglecting any dependance on contamination or incompleteness, i.e. $\mu_{C1}=\mu_{I1}=\mu_{C2}=\mu_{I2}=0$. 
The model corrected for imperfect membership fits the mass biases much better. Even with the 6 extra parameters, it results in 
a much smaller Bayes Information Criterion, with $\Delta\rm BIC \ll -10$ for all methods, indicating very strong evidence for this more complex 
model.

\begin{figure}
\begin{center}
    \leavevmode
    \epsfxsize=8.5cm
    \epsfbox[70 50 480 430]{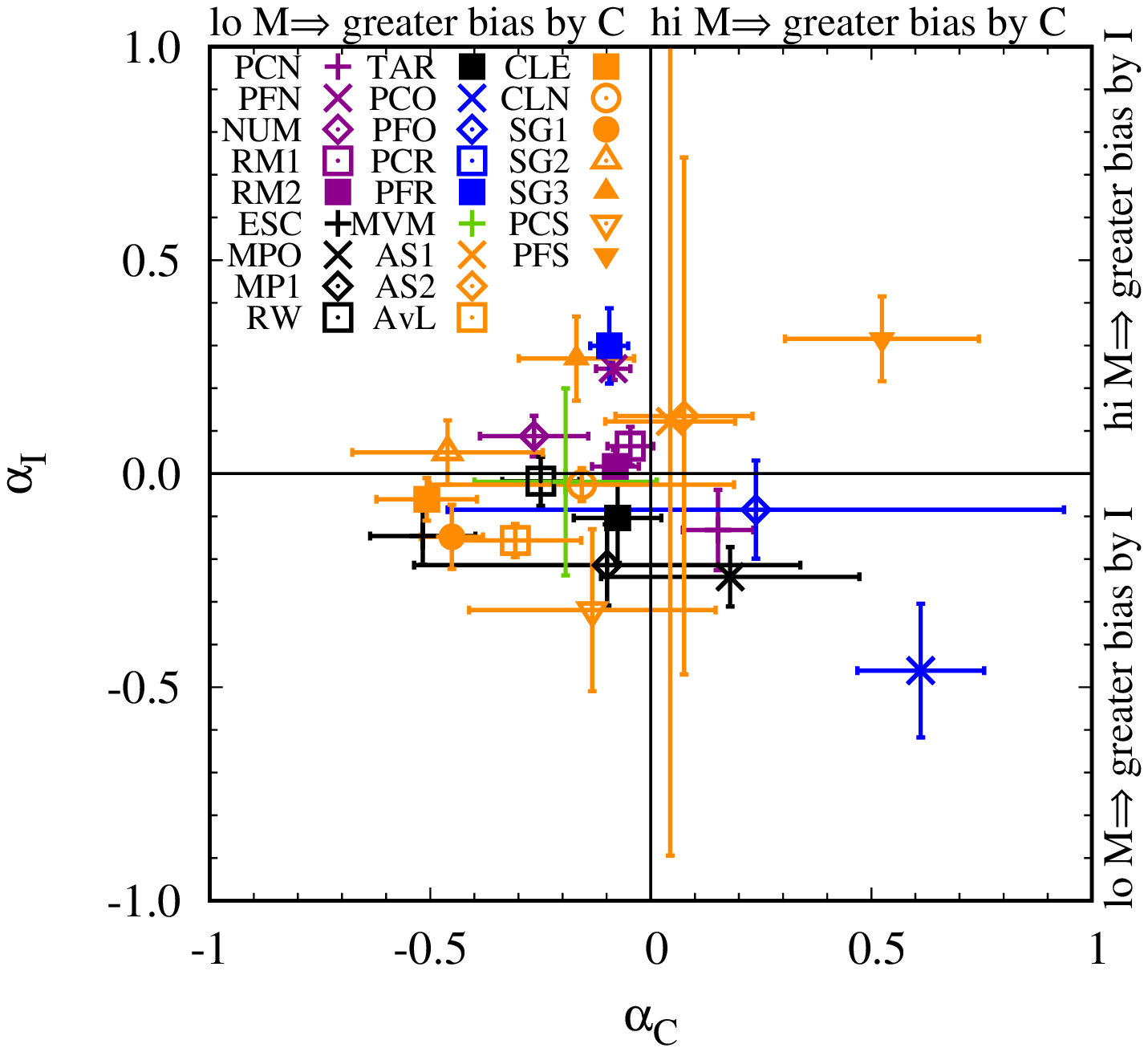}
\end{center}
\caption{Effects of true cluster mass on the variation of mass bias with 
  contamination and incompleteness.
  The mass-dependance is assumed to be a power-law function of true
  cluster mass with slopes $\alpha_{\rm C}$ and $\alpha_{\rm I}$ 
  corresponding to the effect of contamination and incompleteness,
  respectively  (eq.~[\ref{mu_model_g}]). There is a clear of excess methods for which cluster mass
  overestimation due to contamination tend to be stronger for low-mass (lo M) 
  clusters. Method PCR lies out of the bounds of the plot with $\alpha_{\rm C}=-0.7\pm0.1$ 
  and $\alpha_{\rm I}=-1.5\pm0.3$.
 }
\label{CI-mass-dep}
\end{figure}

Fig.~\ref{CI-mass-dep} shows the constraints on the mass-dependent effects of
contamination and incompleteness,
$\alpha_{\rm C}$ and $\alpha_{\rm I}$ respectively, on mass bias, as
determined for each method using equations (\ref{like1}), (\ref{scatter}),
and (\ref{mu_model_g}).  We find that there is a clear excess of 
methods with $\alpha_{\rm C}<0$, i.e. methods for which cluster mass
overestimation due to contamination appears to be stronger for less massive
clusters. On the other hand, the distribution of $\alpha_{\rm I}$ does not reveal 
any significant asymmetry and thus any generic trend in mass dependence of 
bias due to incompleteness.


\begin{figure}
\begin{center}
    \leavevmode
    \epsfxsize=8.5cm
    \epsfbox[70 50 480 430]{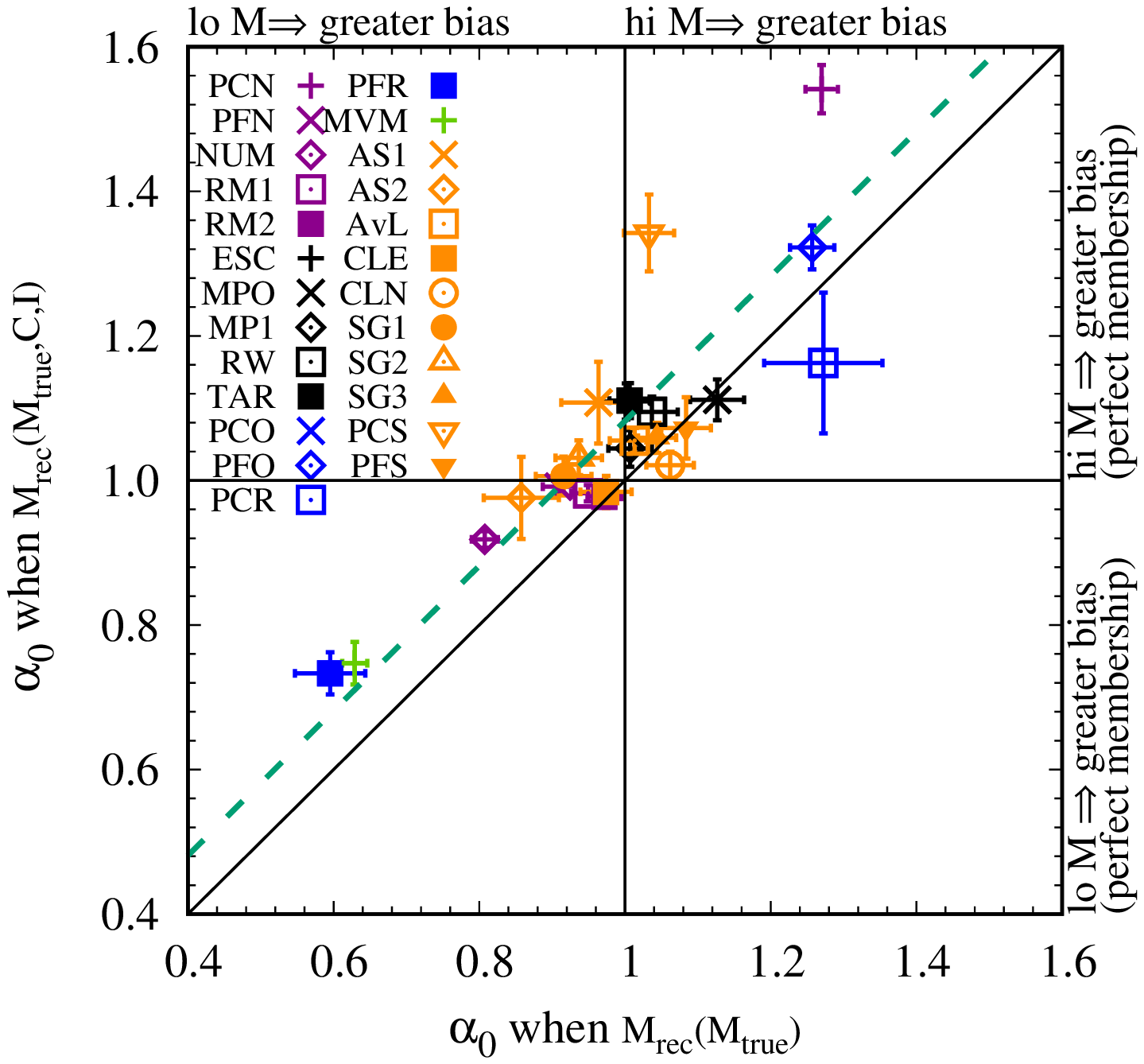}
\end{center}
\caption{Effects of imperfect membership on the slope of the mass scaling
  relation $M_{\rm rec}\sim M_{\rm true}^{\alpha_{0}}$.  The slope is
  measured for each method in two cases: neglecting any prior information on
  membership (\emph{horizontal axis}) and including our full model accounting for
  effects of contamination and incompleteness on cluster mass estimates
  (eq.~[\ref{mu_model_g}], \emph{vertical axis}). The dashed line shows a
  best fitting
  linear model which
  signifies that imperfect membership tends to flatten the $M_{\rm
    rec}-M_{\rm true}$ relation (greater bias for low-mass -- `lo M' --  clusters) 
    reducing its slope on average by 0.084.
}
\label{alpha0}
\end{figure}

\subsection{Effects of imperfect membership on the slope of mass scaling relations}
The slope of the $M_{\rm rec}-M_{\rm true}$ scaling relation remains
independent of contamination and incompleteness if both the selection of cluster
members and the response of the underlying mass estimator to imperfect membership
are independent of cluster mass. Our study demonstrates that this condition
is not satisfied in general. Therefore, one may expect that the effective
slope of the mass scaling relations can be modified to some extent by how
different methods select cluster members. We quantify this effect by
 comparing the slopes measured in two modes: ignoring any dependance of
$\log_{10}(M_{\rm rec}/M_{\rm true})$ on membership
($\mu_{C1}=\mu_{C2}=\mu_{I1}=\mu_{I2}=0$ and free $\alpha_{0}$, $\sigma_{0}$, $\sigma_{1}$) 
and including a complete (generalized) model describing effects of imperfect membership, as given by
equation~(\ref{mu_model_g}). The effective slopes measured in the first mode
\citep[the same as measured and discussed in][]{Old2015} depend both on
intrinsic properties of the mass estimators and imperfect membership
(contamination and incompleteness) of selected galaxy samples, while the
slopes measured in the second mode reflect primarily performance of the mass
estimators. Therefore, comparing the slopes from the two modes is expected to
extract a genuine effect of imperfect membership on the slopes of the mass
scaling relations.

\begin{figure}
\begin{center}
    \leavevmode
    \epsfxsize=8.5cm
    \epsfbox[70 60 480 805]{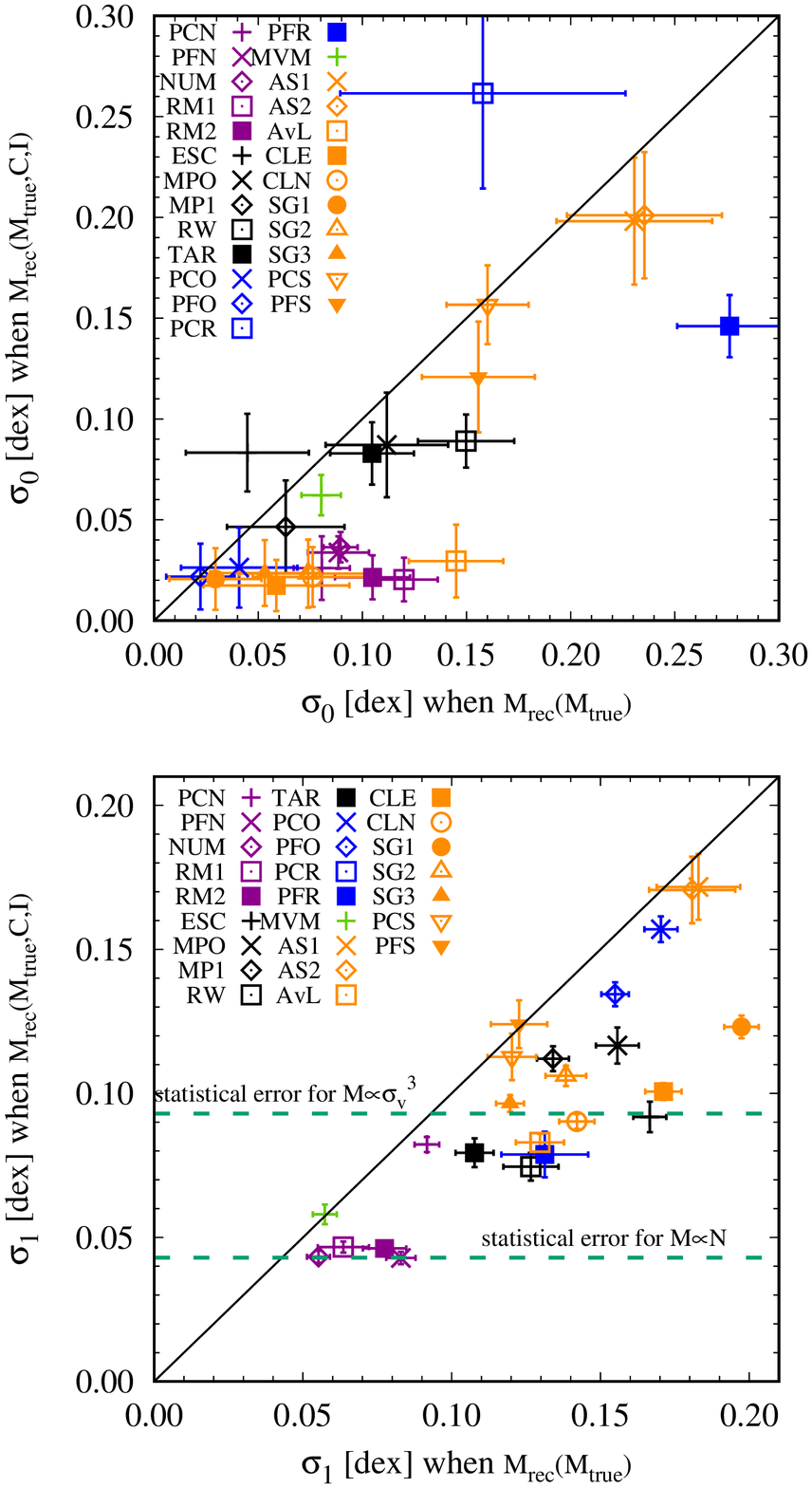}
\end{center}
\caption{Effects of imperfect membership on the Poisson-like scatter
  (\emph{top panel}) and the irreducible scatter (\emph{bottom panel}), as defined by equation
  (\ref{scatter}), in the $M_{\rm rec}-M_{\rm true}$ scaling relations. The
  two scatter parameters are measured in two modes: neglecting any prior
  information on membership (\emph{horizontal axis}) and including a full model
  accounting for effects of contamination and incompleteness on the cluster
  mass estimates (\emph{vertical axis}). The \emph{dashed lines} in the bottom panel
  indicate two characteristic levels of the Poisson-like scatter for
  richness-based and velocity dispersion-based methods. The results
  demonstrate that imperfect membership increases substantially both the
  intrinsic and Poisson-like scatter.}
\label{scatter-2}
\end{figure}

Fig.~\ref{alpha0} shows the constraints on slope $\alpha_{0}$ measured in
the two modes described above. The results clearly demonstrate that imperfect
membership gives rise to a flattening of the $M_{\rm rec}-M_{\rm true}$ scaling
relations for most methods.  We find that the relationship between the two
slopes is well approximated by a linear model with slope $1.01\pm0.11$,
intercept $0.084\pm0.019$ and scatter $0.081\pm0.017$ (see the dashed green line). 
This simple model implies that a typical reduction of the slope
$\alpha_{0}$ due to imperfect membership is equal to $0.084$ and is
independent of the slope found by neglecting the effects of imperfect membership.

\subsection{Effects of imperfect membership on the scatter of mass scaling relations}
\label{sec:scatterM}

We follow the  same approach as
described in the previous subsection to study the effects of imperfect
membership on the scatter in the $M_{\rm
  rec}-M_{\rm true}$ scaling relations. We measure both free parameters of the
effective scatter given by equation~(\ref{scatter}) in two modes: ignoring any
dependance on membership ($\mu_{C1}=\mu_{C2}=\mu_{I1}=\mu_{I2}=0$ and 
free $\alpha_{0}$, $\sigma_{0}$, $\sigma_{1}$) and including a complete (generalized) model describing effects of imperfect
membership, as described by (\ref{mu_model_g}).  Comparing values of
both $\sigma_0$ and $\sigma_1$  determined in these two modes  quantifies the contribution of
imperfect membership to the total scatter in the mass scaling relations.

As demonstrated in Fig.~\ref{scatter-2}, imperfect membership increases 
both the intrinsic and Poisson-like scatter substantially. Applying
corrections due to imperfect membership can potentially bring down the
Poisson-like scatter to theoretically expected levels. This is illustrated in
the bottom panel, which shows theoretical lower limits of the
Poisson-like scatter for two basic mass estimators based on richness or 
velocity dispersion (magenta and orange symbols, respectively), i.e.  $M_{\rm true}\propto N_{\rm mem}$ and $M_{\rm
  true}\propto \sigma_{v}^{3}$. The former is the prediction for
richness-based methods, assuming $\alpha_0=1$ as is roughly the case for most
methods, as seen in Fig.~\ref{alpha0}, leading to $\sigma_1 = 1/(10\,\ln
10)\simeq 0.043$.
The latter is the prediction for methods 
utilizing kinematics (velocity dispersion or distribution in projected phase-space), and
given that the uncertainty on the standard deviation for a Gaussian
distribution is $\epsilon(\sigma) = \sigma/\sqrt{2N}$, the scatter on $\mu
\propto \log_{10} \sigma_v^3$ is $3/(\sqrt{2N}\,\ln 10)$, leading to
$\sigma_1 = 0.3/(\sqrt{2N}\,\ln 10) \simeq 0.092$.
This assumes that any velocity bias of galaxies tracing the gravitational
potential is independent of mass (as found by \citealp{Munari+13}, but
disputed by \citealp*{Old+13}).
All richness-based methods except PCN have their Poisson scatter term, $\sigma_1$, as low as
the theoretical limit of $0.1/(\ln 10)$. Nearly all the velocity dispersion and projected phase
space methods (except AS1 and AS2; orange and black symbols, respectively) have their Poisson scatter term fairly
close to the theoretical limit of $0.3/(\sqrt{2N}\,\ln 10)$.

Imperfect membership also appears to be a primary source of the intrinsic (irreducible)
scatter, $\sigma_0$. For 12 methods, the intrinsic scatter becomes negligible
($\sigma_0 < 0.03$) when mass
estimates are corrected for imperfect membership. But several methods  (AS1, AS2,
PCR, PCS, PFR and PFS) have an important intrinsic scatter ($\sigma_0>0.1$)
after correction
for imperfect membership. Except for PCS, these methods appear to be
substantially dominated by intrinsic errors of the mass estimators and are
amongst those with the lowest ranks of the mass recovery accuracy 
and the rms difference between the recovered and true log cluster
mass larger than 0.45 dex, compared to the rms of $\sim0.20$ dex for the best methods 
\citep[see Table 2 in][]{Old2015}.

\section{Summary and conclusions}

We used mock observations of galaxy clusters in the optical band (photometric
and spectroscopic data) to study the impact of imperfect membership in
selected galaxy samples, as quantified by contamination and incompleteness,
on cluster mass estimates based on 25 different methods employing various
techniques of galaxy selection and dynamical mass estimation. The mock
catalogue was generated using an HOD approach with some improvements to
emulate more realistically observations of galaxy clusters in terms of the colour
distribution of galaxies, the phase-space distribution of satellites,
miscentering of brightest cluster galaxies and several other properties
(see \citealp{Old2015}). Although each of the 25 methods
considered in this study has its own peculiarities, the methods can be
grouped into four broad categories with respect to what part of the available
data is utilized to select cluster members and estimate cluster masses:
methods based on richness, galaxy positions, velocity dispersion, and
projected phase-space distributions.

We demonstrated that contamination and incompleteness give rise to
respectively overestimation and underestimation of the measured cluster
masses. This general rule holds for nearly all methods and for all four
categories of mass estimators.  For each method, it is possible to find a
combination of contamination and incompleteness for which the mass
overestimation due to contamination is fully compensated by its counterpart
due to incompleteness (green lines in Fig.~\ref{CI-HOD2}).
The mass estimation accuracy given by $\langle \log_{10}(M_{\rm rec}/M_{\rm
  true})\rangle$
evaluated for the mean
contamination and incompleteness does not vanish and thus sets an irreducible bias
for each mass estimator. Using a linear model for describing the impact of
imperfect membership about the mean contamination and incompleteness, we find
that all methods exhibit (Fig.~\ref{lin_slopes}) the same dependence on contamination with
$\log_{10}(M_{\rm rec}/M_{\rm true})\sim (0.50\pm0.12)C$, consistent with
$M_{\rm rec}\sim (1+C)$ expected for a simple, idealized richness-based method
with
$M_{\rm  rec}\sim N_{\rm mem}$. On the other hand, the analogous dependance on
incompleteness is not universal, with $\log_{10}(M_{\rm rec}/M_{\rm true})$
in the range of $-2.0\,I$ to $-0.5\,I$, with the highest sensitivity
to incompleteness for methods based on kinematics (velocity dispersion and projected phase-space
analysis).

Imperfect membership modifies the $M_{\rm rec}-M_{\rm true}$ scaling relation
in two respects. The primary effect, demonstrated also in other studies \citep[see e.g.][]{Sar2013}, 
is an increase of scatter (Fig.~\ref{scatter-2}). 
We found that this affects both the intrinsic scatter
and the Poisson-like scatter scaling with $1/N_{\rm mem}^{1/2}$, where
$N_{\rm mem}$ is the number of true cluster members. Secondly, due to a
mass-dependent selection of cluster members (Fig.~\ref{cpdf}) and a mass-dependent response of
the cluster mass estimators to imperfect membership (Fig.~\ref{CI-mass-dep}),
the $M_{\rm rec}-M_{\rm true}$ relation becomes flatter than that based on the assumption of fully
self-similar samples of galaxies selected as cluster members
(Fig.~\ref{alpha0}).
  This flattening
arises from both a higher contamination of galaxy samples
(Fig.~\ref{cpdf})
and a higher
sensitivity of cluster mass estimators to contamination
(Fig.~\ref{CI-differences})
for less massive
systems.

\begin{figure}
\begin{center}
    \leavevmode
    \epsfxsize=8.5cm
    \epsfbox[55 50 475 430]{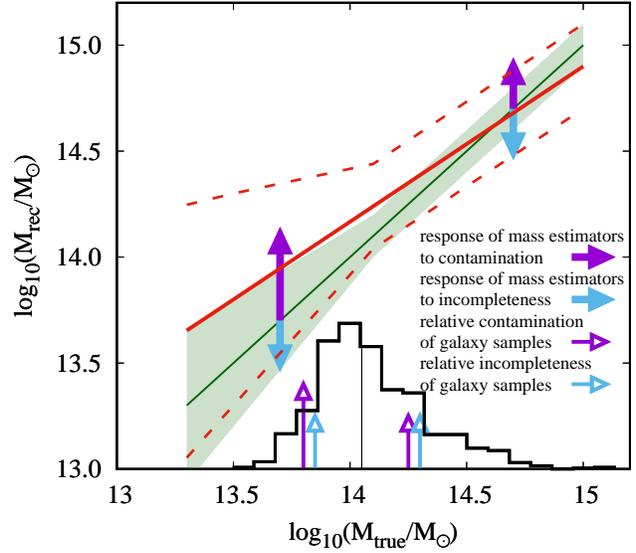}
\end{center}
\caption{Illustration of the effects of imperfect membership on the mass
  scaling relation. The \emph{shaded region} represents the scaling relation
  unaffected by imperfect membership, with the width rendering scatter
  dominated by shot noise at low masses and intrinsic scatter at high
  masses. The \emph{thick solid} and \emph{dashed lines} show the alteration of the
  relation due to typical contamination and incompleteness of selected galaxy
  samples. Imperfect membership flattens the relation and increases both
  types of scatter. The \emph{arrows} show the relative impact of two main effects:
  the response of the mass estimator to imperfect membership and the typical levels of
  contamination/incompleteness in selected galaxy samples. The apparent
  asymmetry of the two effects at different mass scales is responsible for
  flattening of the $M_{\rm rec}-M_{\rm true}$ relation. The \emph{black histogram} shows
  the distribution of true cluster masses in the catalogue, with the median
  mass indicated by the \emph{thin black vertical line}.
}
\label{schematic}
\end{figure}

Fig.~\ref{schematic} schematically illustrates how imperfect membership
affects the $M_{\rm rec}-M_{\rm true}$ scaling relation.  The quantitative
description of effects of imperfect membership on the mass scaling relations
is based on a specific choice of the prior distribution of cluster masses
with a complete sampling of the underlying mass function at $\log_{10}M_{\rm
  true}\gtrsim 14$ and a smooth cut-off at low masses (see the histogram in
Fig.~\ref{schematic}).  We expect that exact values of parameters describing
the $M_{\rm rec}-M_{\rm true}$ relation and their response to imperfect
membership may change when providing cluster samples which are more complete
at lower masses.  Including more systems at low masses, however, can only
enhance the effects illustrated in Fig.~\ref{schematic}, in particular
the flattening of the mass scaling relation and the increase of the Poisson-like scatter.

Our results show that improvement in assigning cluster membership can
substantially reduce the scatter in the mass scaling relations for virtually
all methods. As demonstrated in Fig.~\ref{scatter-2}, accounting for
imperfect membership turns the intrinsic scatter into a subdominant
contribution for most techniques. The exact values of the intrinsic scatter
may be underestimated due to some simplifying assumptions adopted in the HOD
approach to generating galaxy catalogues.  We expect that an additional
contribution to the intrinsic scatter may arise from effects which are not
accounted for in the HOD catalogue, e.g. intrinsic shapes of
galaxy clusters, the orbital anisotropy or realistic substructure. The impact of these effects on the
effective scatter can be quantified using mock observations based on more
realistic models describing connections between galaxy properties and the
underlying dark matter haloes. For example, \citet{Old2018} utilized mock
observations \citep[a part of the GCMRP;][]{Old2015} generated with the
Semi-Analytic Galaxy Evolution (SAGE) galaxy formation model \citep{Cro2006}
and measured an average difference of $0.054$~dex between log masses of
galaxy clusters with or without substructures. Assuming a Gaussian
distribution of their effect on mass estimates, dynamical substructures are
then a source of a $0.03$~dex intrinsic scatter in the mass scaling relations
based on galaxies.

The mass scaling relations studied in our work are of particular importance
for inferring cosmological constraints from the cluster abundance measured in
upcoming cosmological surveys such as Euclid or LSST. From this point of view, our
results shed light on some aspects of observational strategies for cluster cosmology. 
First of all, imperfect membership appears to be the main and for some methods 
the only source of intrinsic scatter in the mass scaling relation. Therefore, it is clear
that further development of most methods for cluster mass estimation shall 
prioritize improvement of algorithms assigning cluster membership. Secondly,
our study shows that the effective slope of the mass scaling relation does
not only reflect the intrinsic performance of the mass estimator, but is also
affected by mass-dependent effects of imperfect membership. Bearing this in
mind, it becomes clear that a self-consistent comparison between observational 
and simulation-based mass scaling relations requires that both mock and real 
cluster data are nearly the same in terms of the mass range and they are analyzed 
using the same algorithms for assigning membership and estimating cluster masses. 

\begin{figure}
\begin{center}
    \leavevmode
    \epsfxsize=8.5cm
    \epsfbox[55 50 475 430]{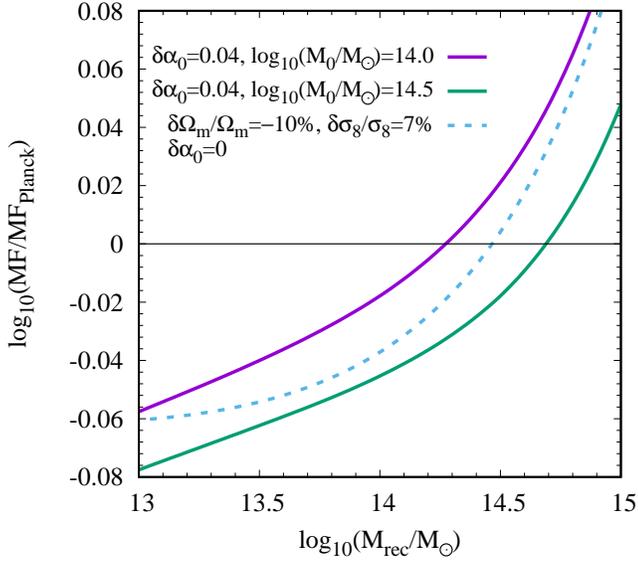}
\end{center}
\caption{The impact of neglecting the flattening of the $M_{\rm rec}-M_{\rm true}$ relation 
due to mass-dependent imperfect membership on the observationally measured mass function (MF). 
It is assumed that the $M_{\rm rec}\propto M_{\rm true}^{\alpha_{0}}$ relation is determined in a very narrow 
mass range about $M_{0}$, leading to a bias $\delta\alpha_{0}$ when extrapolating the relation to lower 
and higher masses. The \emph{dashed line} shows the mass function for modifications of cosmological parameters 
which approximately compensate the effects of mass-dependent imperfect membership. The mass functions 
are calculated relative to a fiducial model given by the Planck cosmology (MF$_{\rm Planck}$).
}
\label{cosmology}
\end{figure}

Uncorrected flattening of the $M_{\rm rec}-M_{\rm true}$ relation 
due to mass-dependent effects of imperfect membership may distort the mass function reproduced 
from observations and consequently bias the measurement of cosmological parameters. Fig.~\ref{cosmology} 
shows the magnitude of this effect in a simplified scenario where we assume that 
the $M_{\rm rec}-M_{\rm true}$ relation is determined in a very narrow mass range about 
$M_{0}$ and its extrapolation to low/high masses results in adopting too high a slope for the actual 
$M_{\rm rec}-M_{\rm true}$ relation valid in the wider mass range. Using only a half of the mean 
increase of $\alpha_{0}$ shown in Fig.~\ref{alpha0}, we find that the observationally reconstructed 
mass function is suppressed at low masses and amplified at high masses by $\sim0.05$~dex ($12$ per cent) 
compared to the fiducial model given by a fitting function from \citet{Tin2008} 
and the Planck cosmology \citep{Planck2016}. As shown by the dashed line, 
this in turn can degrade the accuracy of cosmological inference with $\Omega_{\rm m}$ 
biased down by $\sim10$ per cent  and $\sigma_{8}$  biased up by $\sim 7$ per cent. 
Interestingly, a comparable bias due to unaccounted projection effects for 
richness-based methods to measure cosmological parameters from the cluster mass function 
was demonstrated by \citet{Cos2018}. We think that our estimated biases shall be regarded as upper limits, 
since realistic calibrations of the $M_{\rm rec}-M_{\rm true}$ relation can 
be performed in a mass range comparable to that used 
for measuring the mass function, alleviating the effects of mass-dependent imperfect membership. 
However, the magnitude of the obtained biases in cosmological parameters demonstrates 
the importance of accounting for imperfect membership on all mass scales in a robust way, 
especially when cluster cosmology from future surveys such as Euclid is expected to reach a sub-percent 
precision \citep{Sar2016}. Another complication may arise from a possible redshift dependance 
of imperfect membership. This is not considered in our work, but it is definitely worth studying using 
high-redshift mock observations of galaxy clusters. 

Although our study is solely based on the HOD mock galaxy catalogue of the GCMRP, we confirm 
that all main effects unveiled by the HOD data are also readily visible in the twin galaxy catalogue produced 
by the SAM model \citep{Old2014}. In particular, we find that the $M_{\rm rec}\propto M_{\rm true}^{\alpha_{0}}$ relation 
is affected by a comparable flattening due to a net effect of mass-dependent galaxy selection and sensitivity 
of the mass estimators to imperfect membership, with a mean reduction of $0.075\pm0.032$ in $\alpha_{0}$. 
Consistency between results obtained from the two mock galaxy catalogues as well as the fact 
of using a wide variety of cluster mass reconstruction methods corroborate the generality of the conclusions 
reached in this study.

\section*{Acknowledgments}
We greatly acknowledge the anonymous referee for numerous comments that improved our paper. 
We thank the Instituto de Fisica Teorica (IFT-UAM/CSIC in Madrid) 
and the Universidad Autonoma de Madrid (UAM) for their hospitality and support, via the Centro de Excelencia Severo Ochoa Program under Grant No. 
SEV-2012-0249, during the three week workshop ``nIFTy Cosmology'' where a part of this work developed. RAS acknowledges support from the NSF 
grant AST-1055081. CS acknowledges support from the European Research Council under FP7 grant number 279396. RW was supported by a grant 
from VILLUM FONDEN (Project No. 16599). AS is supported by the ERC-StG ``ClustersXCosmo'', grant agreement 71676. 
ET was supported by ETAg grants IUT40-2, IUT26-2 and by EU through the ERDF CoE grant TK133 and MOBTP86. TS was supported 
by European Regional Development Fund under the grant no EU48684.

\bibliography{master}

\begin{thebibliography}{60}
\expandafter\ifx\csname natexlab\endcsname\relax\def\natexlab#1{#1}\fi

\bibitem[{{Behroozi} {et~al}\mbox{.}(2013){Behroozi}, {Wechsler}, \&
  {Wu}}]{Beh2013}
{Behroozi} P.~S., {Wechsler} R.~H., {Wu} H.-Y., 2013, \apj, 762, 109

\bibitem[{{Bernyk} {et~al}\mbox{.}(2016){Bernyk}, {Croton}, {Tonini},
  {Hodkinson}, {Hassan}, {Garel}, {Duffy}, {Mutch}, {Poole}, \&
  {Hegarty}}]{Ber2016}
{Bernyk} M. {et~al.}, 2016, \apjs, 223, 9

\bibitem[{{Bocquet} {et~al}\mbox{.}(2015){Bocquet}, {Saro}, {Mohr}, {Aird},
  {Ashby}, {Bautz}, {Bayliss}, {Bazin}, {Benson}, {Bleem}, {Brodwin},
  {Carlstrom}, {Chang}, {Chiu}, {Cho}, {Clocchiatti}, {Crawford}, {Crites},
  {Desai}, {de Haan}, {Dietrich}, {Dobbs}, {Foley}, {Forman}, {Gangkofner},
  {George}, {Gladders}, {Gonzalez}, {Halverson}, {Hennig}, {Hlavacek-Larrondo},
  {Holder}, {Holzapfel}, {Hrubes}, {Jones}, {Keisler}, {Knox}, {Lee}, {Leitch},
  {Liu}, {Lueker}, {Luong-Van}, {Marrone}, {McDonald}, {McMahon}, {Meyer},
  {Mocanu}, {Murray}, {Padin}, {Pryke}, {Reichardt}, {Rest}, {Ruel}, {Ruhl},
  {Saliwanchik}, {Sayre}, {Schaffer}, {Shirokoff}, {Spieler}, {Stalder},
  {Stanford}, {Staniszewski}, {Stark}, {Story}, {Stubbs}, {Vanderlinde},
  {Vieira}, {Vikhlinin}, {Williamson}, {Zahn}, \& {Zenteno}}]{Bocq2015}
{Bocquet} S. {et~al.}, 2015, \apj, 799, 214

\bibitem[{{Cataneo} {et~al}\mbox{.}(2015){Cataneo}, {Rapetti}, {Schmidt},
  {Mantz}, {Allen}, {Applegate}, {Kelly}, {von der Linden}, \&
  {Morris}}]{Cat2015}
{Cataneo} M. {et~al.}, 2015, \prd, 92, 044009

\bibitem[{{Cava} {et~al}\mbox{.}(2017){Cava}, {Biviano}, {Mamon}, {Varela},
  {Bettoni}, {D'Onofrio}, {Fasano}, {Fritz}, {Moles}, {Moretti}, \&
  {Poggianti}}]{Cav2017}
{Cava} A. {et~al.}, 2017, \aap, 606, A108

\bibitem[{{Collister} \& {Lahav}(2005)}]{Coll2005}
{Collister} A.~A., {Lahav} O., 2005, \mnras, 361, 415

\bibitem[{{Costanzi} {et~al}\mbox{.}(2018){Costanzi}, {Rozo}, {Rykoff},
  {Farahi}, {Jeltema}, {Evrard}, {Mantz}, {Gruen}, {Mandelbaum}, {DeRose},
  {McClintock}, {Varga}, {Zhang}, {Weller}, {Wechsler}, \& {Aguena}}]{Cos2018}
{Costanzi} M. {et~al.}, 2018, ArXiv 1807.07072

\bibitem[{{Croton} {et~al}\mbox{.}(2006){Croton}, {Springel}, {White}, {De
  Lucia}, {Frenk}, {Gao}, {Jenkins}, {Kauffmann}, {Navarro}, \&
  {Yoshida}}]{Cro2006}
{Croton} D.~J. {et~al.}, 2006, \mnras, 365, 11

\bibitem[{{de Carvalho} {et~al}\mbox{.}(2017){de Carvalho}, {Ribeiro},
  {Stalder}, {Rosa}, {Costa}, \& {Moura}}]{Car2017}
{de Carvalho} R.~R., {Ribeiro} A.~L.~B., {Stalder} D.~H., {Rosa} R.~R., {Costa}
  A.~P., {Moura} T.~C., 2017, \aj, 154, 96

\bibitem[{{den Hartog} \& {Katgert}(1996)}]{Har1996}
{den Hartog} R., {Katgert} P., 1996, \mnras, 279, 349

\bibitem[{{Diaferio}(1999)}]{Dia99}
{Diaferio} A., 1999, \mnras, 309, 610

\bibitem[{{Duarte} \& {Mamon}(2014)}]{Duarte&Mamon14}
{Duarte} M., {Mamon} G.~A., 2014, \mnras, 440, 1763

\bibitem[{{Duarte} \& {Mamon}(2015)}]{Duarte&Mamon15}
{Duarte} M., {Mamon} G.~A., 2015, \mnras, 453, 3848

\bibitem[{{Fadda} {et~al}\mbox{.}(1996){Fadda}, {Girardi}, {Giuricin},
  {Mardirossian}, \& {Mezzetti}}]{Fad1996}
{Fadda} D., {Girardi} M., {Giuricin} G., {Mardirossian} F., {Mezzetti} M.,
  1996, \apj, 473, 670

\bibitem[{{Gifford} \& {Miller}(2013)}]{Gif2013}
{Gifford} D., {Miller} C.~J., 2013, \apjl, 768, L32

\bibitem[{{Gladders} \& {Yee}(2000)}]{Glad2000}
{Gladders} M.~D., {Yee} H.~K.~C., 2000, \aj, 120, 2148

\bibitem[{{Jimeno} {et~al}\mbox{.}(2017){Jimeno}, {Broadhurst}, {Lazkoz},
  {Angulo}, {Diego}, {Umetsu}, \& {Chu}}]{Jim2017}
{Jimeno} P., {Broadhurst} T., {Lazkoz} R., {Angulo} R., {Diego} J.-M., {Umetsu}
  K., {Chu} M.-c., 2017, \mnras, 466, 2658

\bibitem[{{Klypin} {et~al}\mbox{.}(2011){Klypin}, {Trujillo-Gomez}, \&
  {Primack}}]{Kly11}
{Klypin} A.~A., {Trujillo-Gomez} S., {Primack} J., 2011, \apj, 740, 102

\bibitem[{{Knebe} {et~al}\mbox{.}(2011){Knebe}, {Knollmann}, {Muldrew},
  {Pearce}, {Aragon-Calvo}, {Ascasibar}, {Behroozi}, {Ceverino}, {Colombi},
  {Diemand}, {Dolag}, {Falck}, {Fasel}, {Gardner}, {Gottl{\"o}ber}, {Hsu},
  {Iannuzzi}, {Klypin}, {Luki{\'c}}, {Maciejewski}, {McBride}, {Neyrinck},
  {Planelles}, {Potter}, {Quilis}, {Rasera}, {Read}, {Ricker}, {Roy},
  {Springel}, {Stadel}, {Stinson}, {Sutter}, {Turchaninov}, {Tweed}, {Yepes},
  \& {Zemp}}]{Kne2011}
{Knebe} A. {et~al.}, 2011, \mnras, 415, 2293

\bibitem[{{Lopes} {et~al}\mbox{.}(2009){Lopes}, {de Carvalho}, {Kohl-Moreira},
  \& {Jones}}]{Lopes2009}
{Lopes} P.~A.~A., {de Carvalho} R.~R., {Kohl-Moreira} J.~L., {Jones} C., 2009,
  \mnras, 392, 135

\bibitem[{{Mamon} {et~al}\mbox{.}(2013){Mamon}, {Biviano}, \&
  {Bou{\'e}}}]{Mam2013}
{Mamon} G.~A., {Biviano} A., {Bou{\'e}} G., 2013, \mnras, 429, 3079

\bibitem[{{Mamon} {et~al}\mbox{.}(2010){Mamon}, {Biviano}, \&
  {Murante}}]{Mam2010}
{Mamon} G.~A., {Biviano} A., {Murante} G., 2010, \aap, 520, A30

\bibitem[{{Mantz} {et~al}\mbox{.}(2014){Mantz}, {Allen}, {Morris}, {Rapetti},
  {Applegate}, {Kelly}, {von der Linden}, \& {Schmidt}}]{Man2014}
{Mantz} A.~B., {Allen} S.~W., {Morris} R.~G., {Rapetti} D.~A., {Applegate}
  D.~E., {Kelly} P.~L., {von der Linden} A., {Schmidt} R.~W., 2014, \mnras,
  440, 2077

\bibitem[{{Mantz} {et~al}\mbox{.}(2015){Mantz}, {von der Linden}, {Allen},
  {Applegate}, {Kelly}, {Morris}, {Rapetti}, {Schmidt}, {Adhikari}, {Allen},
  {Burchat}, {Burke}, {Cataneo}, {Donovan}, {Ebeling}, {Shandera}, \&
  {Wright}}]{Man2015}
{Mantz} A.~B. {et~al.}, 2015, \mnras, 446, 2205

\bibitem[{{Mu{\~n}oz-Cuartas} \& {M{\"u}ller}(2012)}]{Mun2012}
{Mu{\~n}oz-Cuartas} J.~C., {M{\"u}ller} V., 2012, \mnras, 423, 1583

\bibitem[{{Munari} {et~al}\mbox{.}(2013){Munari}, {Biviano}, {Borgani},
  {Murante}, \& {Fabjan}}]{Munari+13}
{Munari} E., {Biviano} A., {Borgani} S., {Murante} G., {Fabjan} D., 2013,
  \mnras, 430, 2638

\bibitem[{{Munari} {et~al}\mbox{.}(2014){Munari}, {Biviano}, \&
  {Mamon}}]{Mun2014}
{Munari} E., {Biviano} A., {Mamon} G.~A., 2014, \aap, 566, A68

\bibitem[{{Ntampaka} {et~al}\mbox{.}(2015){Ntampaka}, {Trac}, {Sutherland},
  {Battaglia}, {P{\'o}czos}, \& {Schneider}}]{Nta2015}
{Ntampaka} M., {Trac} H., {Sutherland} D.~J., {Battaglia} N., {P{\'o}czos} B.,
  {Schneider} J., 2015, \apj, 803, 50

\bibitem[{{Ntampaka} {et~al}\mbox{.}(2016){Ntampaka}, {Trac}, {Sutherland},
  {Fromenteau}, {P{\'o}czos}, \& {Schneider}}]{Nta2016}
{Ntampaka} M., {Trac} H., {Sutherland} D.~J., {Fromenteau} S., {P{\'o}czos} B.,
  {Schneider} J., 2016, \apj, 831, 135

\bibitem[{{Old} {et~al}\mbox{.}(2013){Old}, {Gray}, \& {Pearce}}]{Old+13}
{Old} L., {Gray} M.~E., {Pearce} F.~R., 2013, \mnras, 434, 2606

\bibitem[{{Old} {et~al}\mbox{.}(2014){Old}, {Skibba}, {Pearce}, {Croton},
  {Muldrew}, {Mu{\~n}oz-Cuartas}, {Gifford}, {Gray}, {von der Linden}, {Mamon},
  {Merrifield}, {M{\"u}ller}, {Pearson}, {Ponman}, {Saro}, {Sepp}, {Sif{\'o}n},
  {Tempel}, {Tundo}, {Wang}, \& {Wojtak}}]{Old2014}
{Old} L. {et~al.}, 2014, \mnras, 441, 1513

\bibitem[{{Old} {et~al}\mbox{.}(2015){Old}, {Wojtak}, {Mamon}, {Skibba},
  {Pearce}, {Croton}, {Bamford}, {Behroozi}, {de Carvalho},
  {Mu{\~n}oz-Cuartas}, {Gifford}, {Gray}, {der Linden}, {Merrifield},
  {Muldrew}, {M{\"u}ller}, {Pearson}, {Ponman}, {Rozo}, {Rykoff}, {Saro},
  {Sepp}, {Sif{\'o}n}, \& {Tempel}}]{Old2015}
{Old} L. {et~al.}, 2015, \mnras, 449, 1897

\bibitem[{{Old} {et~al}\mbox{.}(2018){Old}, {Wojtak}, {Pearce}, {Gray},
  {Mamon}, {Sif{\'o}n}, {Tempel}, {Biviano}, {Yee}, {de Carvalho},
  {M{\"u}ller}, {Sepp}, {Skibba}, {Croton}, {Bamford}, {Power}, {von der
  Linden}, \& {Saro}}]{Old2018}
{Old} L. {et~al.}, 2018, \mnras, 475, 853

\bibitem[{{Pearson} {et~al}\mbox{.}(2015){Pearson}, {Ponman}, {Norberg},
  {Robotham}, \& {Farr}}]{Pearson2015}
{Pearson} R.~J., {Ponman} T.~J., {Norberg} P., {Robotham} A.~S.~G., {Farr}
  W.~M., 2015, \mnras, 449, 3082

\bibitem[{{Planck Collaboration} {et~al}\mbox{.}(2016){Planck Collaboration},
  {Ade}, {Aghanim}, {Arnaud}, {Ashdown}, {Aumont}, {Baccigalupi}, {Banday},
  {Barreiro}, {Bartlett}, \& et~al.}]{Planck2016}
{Planck Collaboration} {et~al.}, 2016, \aap, 594, A13

\bibitem[{{Roberts} \& {Parker}(2017)}]{Rob2017}
{Roberts} I.~D., {Parker} L.~C., 2017, \mnras, 467, 3268

\bibitem[{{Rozo} {et~al}\mbox{.}(2010){Rozo}, {Wechsler}, {Rykoff}, {Annis},
  {Becker}, {Evrard}, {Frieman}, {Hansen}, {Hao}, {Johnston}, {Koester},
  {McKay}, {Sheldon}, \& {Weinberg}}]{Roz2010}
{Rozo} E. {et~al.}, 2010, \apj, 708, 645

\bibitem[{{Rykoff} {et~al}\mbox{.}(2014){Rykoff}, {Rozo}, {Busha}, {Cunha},
  {Finoguenov}, {Evrard}, {Hao}, {Koester}, {Leauthaud}, {Nord}, {Pierre},
  {Reddick}, {Sadibekova}, {Sheldon}, \& {Wechsler}}]{Ryk2014}
{Rykoff} E.~S. {et~al.}, 2014, \apj, 785, 104

\bibitem[{{Sadeh} {et~al}\mbox{.}(2015){Sadeh}, {Feng}, \& {Lahav}}]{Sad2015}
{Sadeh} I., {Feng} L.~L., {Lahav} O., 2015, Physical Review Letters, 114,
  071103

\bibitem[{{Saro} {et~al}\mbox{.}(2013){Saro}, {Mohr}, {Bazin}, \&
  {Dolag}}]{Sar2013}
{Saro} A., {Mohr} J.~J., {Bazin} G., {Dolag} K., 2013, \apj, 772, 47

\bibitem[{{Sartoris} {et~al}\mbox{.}(2016){Sartoris}, {Biviano}, {Fedeli},
  {Bartlett}, {Borgani}, {Costanzi}, {Giocoli}, {Moscardini}, {Weller},
  {Ascaso}, {Bardelli}, {Maurogordato}, \& {Viana}}]{Sar2016}
{Sartoris} B. {et~al.}, 2016, \mnras, 459, 1764

\bibitem[{{Sif{\'o}n} {et~al}\mbox{.}(2013){Sif{\'o}n}, {Menanteau},
  {Hasselfield}, {Marriage}, {Hughes}, {Barrientos}, {Gonz{\'a}lez}, {Infante},
  {Addison}, {Baker}, {Battaglia}, {Bond}, {Crichton}, {Das}, {Devlin},
  {Dunkley}, {D{\"u}nner}, {Gralla}, {Hajian}, {Hilton}, {Hincks}, {Kosowsky},
  {Marsden}, {Moodley}, {Niemack}, {Nolta}, {Page}, {Partridge}, {Reese},
  {Sehgal}, {Sievers}, {Spergel}, {Staggs}, {Thornton}, {Trac}, \&
  {Wollack}}]{Sifon2013}
{Sif{\'o}n} C. {et~al.}, 2013, \apj, 772, 25

\bibitem[{{Skibba} {et~al}\mbox{.}(2006){Skibba}, {Sheth}, {Connolly}, \&
  {Scranton}}]{Ski2006}
{Skibba} R., {Sheth} R.~K., {Connolly} A.~J., {Scranton} R., 2006, \mnras, 369,
  68

\bibitem[{{Skibba} \& {Sheth}(2009)}]{Ski2009}
{Skibba} R.~A., {Sheth} R.~K., 2009, \mnras, 392, 1080

\bibitem[{{Skibba} {et~al}\mbox{.}(2011){Skibba}, {van den Bosch}, {Yang},
  {More}, {Mo}, \& {Fontanot}}]{Ski2011a}
{Skibba} R.~A., {van den Bosch} F.~C., {Yang} X., {More} S., {Mo} H.,
  {Fontanot} F., 2011, \mnras, 410, 417

\bibitem[{{Skielboe} {et~al}\mbox{.}(2012){Skielboe}, {Wojtak}, {Pedersen},
  {Rozo}, \& {Rykoff}}]{Ski12}
{Skielboe} A., {Wojtak} R., {Pedersen} K., {Rozo} E., {Rykoff} E.~S., 2012,
  \apjl, 758, L16

\bibitem[{{Soares-Santos} {et~al}\mbox{.}(2011){Soares-Santos}, {de Carvalho},
  {Annis}, {Gal}, {La Barbera}, {Lopes}, {Wechsler}, {Busha}, \&
  {Gerke}}]{Soa2011}
{Soares-Santos} M. {et~al.}, 2011, \apj, 727, 45

\bibitem[{{Sunyaev} \& {Zeldovich}(1970)}]{Sunyaev&Zeldovich70}
{Sunyaev} R.~A., {Zeldovich} Y.~B., 1970, \apss, 7, 3

\bibitem[{{Tempel} {et~al}\mbox{.}(2014){Tempel}, {Tamm}, {Gramann},
  {Tuvikene}, {Liivam{\"a}gi}, {Suhhonenko}, {Kipper}, {Einasto}, \&
  {Saar}}]{Tem2014}
{Tempel} E. {et~al.}, 2014, \aap, 566, A1

\bibitem[{{The} \& {White}(1986)}]{The&White86}
{The} L.~S., {White} S.~D.~M., 1986, \aj, 92, 1248

\bibitem[{{Tinker} {et~al}\mbox{.}(2008){Tinker}, {Kravtsov}, {Klypin},
  {Abazajian}, {Warren}, {Yepes}, {Gottl{\"o}ber}, \& {Holz}}]{Tin2008}
{Tinker} J., {Kravtsov} A.~V., {Klypin} A., {Abazajian} K., {Warren} M.,
  {Yepes} G., {Gottl{\"o}ber} S., {Holz} D.~E., 2008, \apj, 688, 709

\bibitem[{{Trevisan} {et~al}\mbox{.}(2017){Trevisan}, {Mamon}, \&
  {Stalder}}]{Trevisan+17}
{Trevisan} M., {Mamon} G.~A., {Stalder} D.~H., 2017, \mnras, 471, L47

\bibitem[{{Von Der Linden} {et~al}\mbox{.}(2007){Von Der Linden}, {Best},
  {Kauffmann}, \& {White}}]{Linden2007}
{Von Der Linden} A., {Best} P.~N., {Kauffmann} G., {White} S.~D.~M., 2007,
  \mnras, 379, 867

\bibitem[{{Wojtak} {et~al}\mbox{.}(2011){Wojtak}, {Hansen}, \&
  {Hjorth}}]{Woj2011}
{Wojtak} R., {Hansen} S.~H., {Hjorth} J., 2011, \nat, 477, 567

\bibitem[{{Wojtak} \& {{\L}okas}(2010)}]{Woj10}
{Wojtak} R., {{\L}okas} E.~L., 2010, \mnras, 408, 2442

\bibitem[{{Wojtak} {et~al}\mbox{.}(2009){Wojtak}, {{\L}okas}, {Mamon}, \&
  {Gottl{\"o}ber}}]{Woj09}
{Wojtak} R., {{\L}okas} E.~L., {Mamon} G.~A., {Gottl{\"o}ber} S., 2009, \mnras,
  399, 812

\bibitem[{{Wojtak} {et~al}\mbox{.}(2007){Wojtak}, {{\L}okas}, {Mamon},
  {Gottl{\"o}ber}, {Prada}, \& {Moles}}]{Woj2007}
{Wojtak} R., {{\L}okas} E.~L., {Mamon} G.~A., {Gottl{\"o}ber} S., {Prada} F.,
  {Moles} M., 2007, \aap, 466, 437

\bibitem[{{Wojtak} \& {Mamon}(2013)}]{Woj2013}
{Wojtak} R., {Mamon} G.~A., 2013, \mnras, 428, 2407

\bibitem[{{Yahil} \& {Vidal}(1977)}]{Yah1977}
{Yahil} A., {Vidal} N.~V., 1977, \apj, 214, 347

\bibitem[{{Yang} {et~al}\mbox{.}(2007){Yang}, {Mo}, {van den Bosch},
  {Pasquali}, {Li}, \& {Barden}}]{Yang2007}
{Yang} X., {Mo} H.~J., {van den Bosch} F.~C., {Pasquali} A., {Li} C., {Barden}
  M., 2007, \apj, 671, 153

\end{thebibliography}

\appendix
\onecolumn

\begin{table*}
\begin{flushleft}
\section{Properties of the Mass Reconstruction Methods}
\label{sec:Methods}
\end{flushleft}
 \centering
  \caption{Detailed description of the member galaxy selection process for all methods}
  \begin{tabular}{p{1.1cm} p{5cm} p{5cm} p{5cm}}
 \toprule
 \multirow{2}{1cm}{\textbf{Methods}}&\multicolumn{3}{c}{Member galaxy selection methodology} \\[1.5ex]
 \cline{2-4}\\[-1.3ex]
 &Initial Galaxy selection&Membership refinement&Treatment of interlopers \\[0.15ex]
 \midrule
 \textcolor{darkmagenta}{\textbf{PCN}}&Within $\rm 5\,Mpc$, $\rm 1000\,km\,s^{-1}$&Clipping of $\pm3\,\sigma$, using galaxies within $\rm 1\,Mpc$&Use galaxies at $\rm 3-5 \,Mpc$ to find interloper population to remove \\
 \textcolor{darkmagenta}{\textbf{PFN}}&Friends of Friends (FoF)&No&No \\
 
 \textcolor{darkmagenta}{\textbf{NUM}}&Within $\rm 3\,Mpc$, $\rm 4000\,km\,s^{-1}$
&1) Estimate $R_{\rm 200c}$  from the relationship between $R_{\rm 200c}$  and
 richness deduced from CLE; 2) Select galaxies within $R_{\rm 200c}$  and with
  $|v|<2.7\,\sigma_{\rm los}^{\rm NFW}(R)$&Same as CLE \\
  \textcolor{darkmagenta}{\textbf{RM1}}&Red sequence&Red sequence& Probabilistic \\
  \textcolor{darkmagenta}{\textbf{RM2}}&Red sequence&Red sequence& Probabilistic \\
 
 \textcolor{black}{\textbf{ESC}}&Within preliminary $R_{\rm 200c}$  estimate and $3500 \,km\,s^{-1}$&Gapper technique&Removed by Gapper technique\\
 
 \textcolor{black}{\textbf{MPO}}&Input from CLN&1) Calculate $R_{\rm 200c}$, $R_{\rm \rho}$, $R_{\rm red}$, $R_{\rm blue}$ by MAMPOSSt method; 2) Select members within radius according to colour&No \\
 
 \textcolor{black}{\textbf{MP1}}&Input from CLN&Same as MPO, except colour blind&No \\
 
 \textcolor{black}{\textbf{RW}}&Within $\rm 3\,\,Mpc$,
 $\rm 4000\,\,km\,s^{-1}$&Within $R_{\rm 200c}$ and $|2\Phi(R)|^{1/2}$, where
 $R_{\rm 200c}$  obtained iteratively &No \\

 \textcolor{black}{\textbf{TAR}}&FoF&No&No \\

 \textcolor{blue}{\textbf{PCO}}&Input from PCN& Input from PCN&Include interloper contamination in density fitting \\
 
 \textcolor{blue}{\textbf{PFO}}&Input from PFN&Input from PFN&No \\
 
 \textcolor{blue}{\textbf{PCR}}&Input from PCN&Input from PCN&Same as PCN \\
 
 \textcolor{blue}{\textbf{PFR}}&Input from PFN&Input from PFN&No \\
 
 \textcolor{green}{\textbf{MVM}}&FoF (ellipsoidal search range, centre of
 most luminous galaxy)&Increasing mass limits, then FoF, loops until closure condition&No \\

 \textcolor{darkorange}{\textbf{AS1}}&Within $\rm 1\,Mpc$, $\rm 4000\,km\,s^{-1}$, constrained by colour-magnitude relation&Clipping of $\pm3\,\sigma$&Removed by clipping of $\pm3\,\sigma$ \\
 
 \textcolor{darkorange}{\textbf{AS2}}&Within $\rm 1\,Mpc$, $4\rm 000\,km\,s^{-1}$, constrained by colour-magnitude relation&Clipping of $\pm3\,\sigma$&Removed by clipping of $\pm3\,\sigma$ \\
 
 \textcolor{darkorange}{\textbf{AvL}}&Within $2.5\,\sigma_{v}$ and $0.8\,R_{\rm 200}$&Obtain $R_{\rm 200c}$  and $\sigma_{v}$ by $\sigma$-clipping&Implicit with $\sigma$-clipping \\
 
 \textcolor{darkorange}{\textbf{CLE}}&Within $\rm 3\,Mpc$, $\rm 4000\,km\,s^{-1}$&1)
 Estimate $R_{\rm 200c}$  from the aperture velocity dispersion; 2) Select galaxies within
 $R_{\rm 200c}$  and with $|v|<2.7\,\sigma_{\rm los}^{\rm NFW}(R)$; 3)
 Iterate steps 1 and 2 until convergence&Obvious interlopers are removed by
 velocity gap technique, then further treated by iterative local $2.7\,\sigma(R)$ clipping \\
 
 \textcolor{darkorange}{\textbf{CLN}}&Input from NUM&Same as CLE&Same as CLE \\
 
 \textcolor{darkorange}{\textbf{SG1}}&Within $\rm 4000 \,km\,s^{-1}$&1) Measure $\sigma_{\rm gal}$, estimate $M_{\rm 200c}$ and $R_{\rm 200c}$; 2) Select galaxies within $R_{\rm 200c}$; 3) Iterate steps 1 and 2 until convergence&Shifting gapper with minimum bin size of $\rm 250 \,kpc$ and 15 galaxies; velocity limit $\rm 1000\,km\,s^{-1}$ from main body \\
 
 \textcolor{darkorange}{\textbf{SG2}}&Within $\rm 4000 \,km\,s^{-1}$&1) Measure $\sigma_{\rm gal}$, estimate $M_{\rm 200c}$ and $R_{\rm 200c}$; 2) Select galaxies within $R_{\rm 200c}$; 3) Iterate steps 1 and 2 until convergence&Shifting gapper with minimum bin size of $\rm 150 \,kpc$ and 10 galaxies; velocity limit $\rm 500 \,km\,s^{-1}$ from main body \\
  \textcolor{darkorange}{\textbf{SG3}}&Within $2.5\,h^{-1}\,\rm Mpc$ and $\rm   4000\,km\,s^{-1}$. Velocity distribution symmeterised& Measure $\sigma_{\rm gal}$, correct for velocity errors, then estimate $M_{\rm 200c}$ and $R_{\rm 200c}$  and apply the
surface pressure term correction (The \& White 1986)&Shifting gapper with
minimum bin size of 
$\rm 420\,h^{-1} kpc$ and 15 galaxies\\
 \textcolor{darkorange}{\textbf{PCS}}&Input from PCN&Input from PCN&Same as PCN \\
 \textcolor{darkorange}{\textbf{PFS}}&Input from PFN&Input from PFN&No \\
  \bottomrule
  \end{tabular}
  \nocite{The&White86}
  \parbox{\hsize}{Notes: The colour of the acronym for each method colour
    corresponds to the main galaxy population property used to perform mass
    estimation richness (magenta), radii
    (blue), velocity dispersion (red), projected phase-space (black), or abundance matching (green). The
    second column details how each method selects an initial member galaxy
    sample, while the third column outlines the member galaxy sample refining
    process. Finally, the fourth column describes how methods treat
    interloping galaxies that are not associated with the clusters.} 
  \label{tab:appendix_table_1}
  
\end{table*}


\begin{figure*}
\section{Goodness of fit}
\label{sec:goodness}
\begin{center}
    \leavevmode
    \epsfxsize=15.8cm
    \epsfbox[55 75 800 1100]{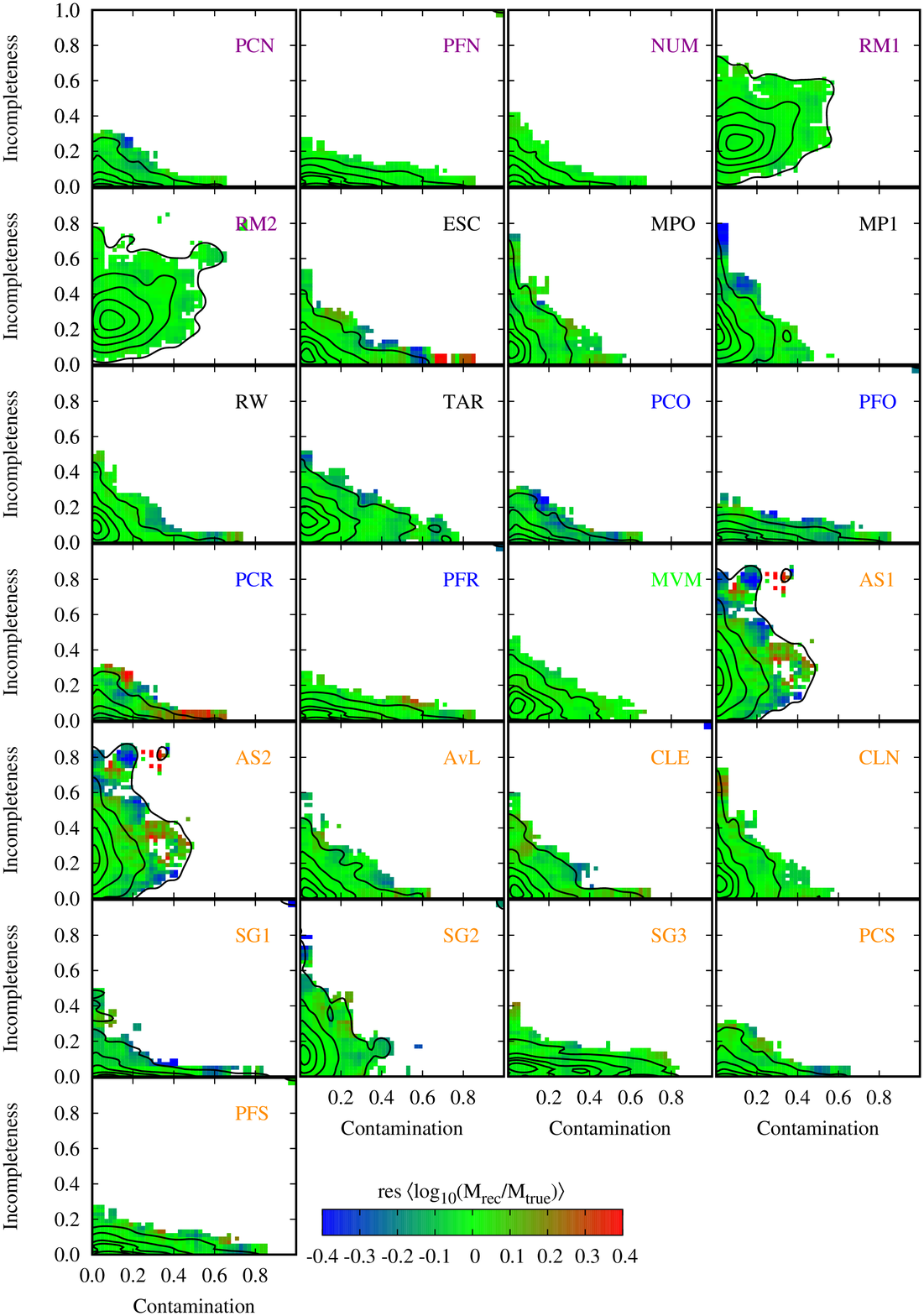}
\end{center}
\caption{Residual maps demonstrating robustness of the best fit models, as given by equation~(\ref{mu_model}), in recovering 
the dependence of the mass estimate accuracy of different methods on contamination and incompleteness of selected 
galaxy samples (see Fig.~\ref{CI-HOD2}).
}
\label{CI-HOD2_res}
\end{figure*}

\begin{figure*}
\section{Galaxy sample selection function}
\label{sec:selection}
\begin{center}
    \leavevmode
    \epsfxsize=15.8cm
    \epsfbox[55 63 800 1100]{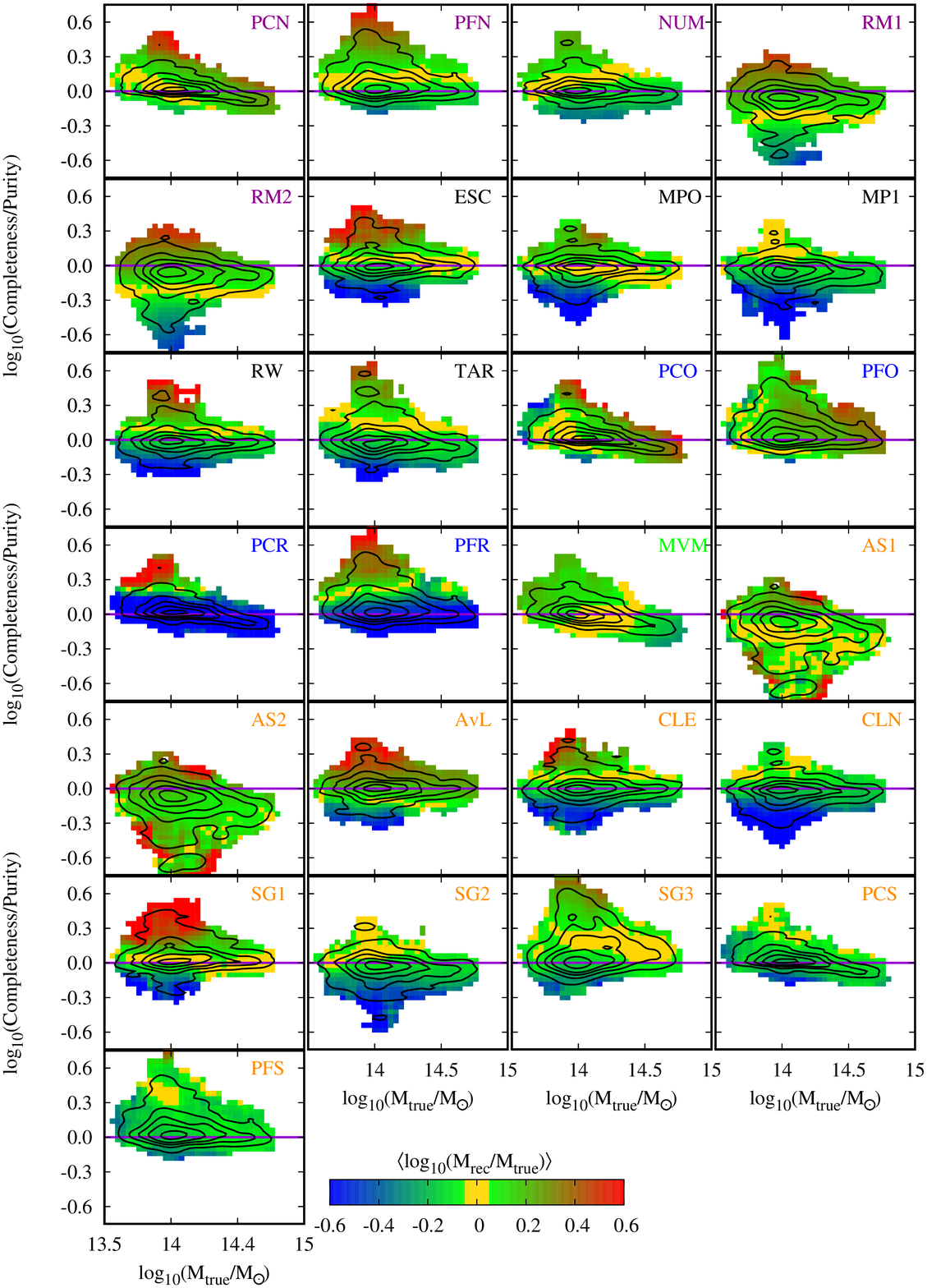}
\end{center}
\caption{Mass dependence of the galaxy sample selection function for all 25 methods of 
cluster mass estimation. The selection function is defined as the ratio of completeness to purity, where 
completeness~$=1-$~incompleteness and purity~$=1-$~contamination. The black contours show the distribution 
of the selection function evaluated at every individual measurement and the true cluster mass. The colour maps show 
the mean mass bias. Both the maps and contours were computed using the same technique as in Fig.~\ref{CI-HOD2}.
}
\label{CP-HOD2}
\end{figure*}


\begin{table*}
\section{Table of results}
\begin{tabular}{llllllllclllllllllll}
  \hline
       &  \multicolumn{6}{c}{Simplistic} & & \multicolumn{8}{c}{Corrected for
        imperfectness} \\
      \cline{2-7}
      \cline{9-16}
      Method &
      $\left\langle C\right\rangle$ &
      $\left\langle I\right\rangle$ &
      $\mu_0$ &
      $\sigma_0$ &
      $\sigma_1$ &
      $\alpha_0$ &
      &
      $\mu_0$ &
      $\mu_{C1}$ &
      $\mu_{C2}$ &
      $\mu_{I1}$ &
      $\mu_{I2}$ &
      $\sigma_0$ &
      $\sigma_1$ &
      $\alpha_0$ \\
      \hline
      
  PCN &   0.13 &   0.06 &   0.10 &   0.08 &   0.09 &   1.27 &  &   0.08 &   0.63 &   0.80 &  -0.65 &  -1.00 &   0.03 &   0.08 &   1.54   \\
  PFN &   0.22 &   0.06 &  -0.01 &   0.09 &   0.08 &   0.91 &  &  -0.02 &   0.61 &   0.51 &  -0.79 &  -0.79 &   0.03 &   0.04 &   0.99   \\
  NUM &   0.11 &   0.06 &  -0.06 &   0.09 &   0.06 &   0.81 &  &  -0.07 &   0.38 &   0.10 &  -1.22 &   1.29 &   0.04 &   0.04 &   0.92   \\
  RM1 &   0.20 &   0.33 &   0.13 &   0.12 &   0.06 &   0.95 &  &   0.14 &   0.58 &   0.36 &  -0.75 &  -0.79 &   0.02 &   0.05 &   0.98   \\
  RM2 &   0.19 &   0.34 &   0.12 &   0.10 &   0.08 &   0.97 &  &   0.13 &   0.57 &   0.50 &  -0.79 &  -0.86 &   0.02 &   0.05 &   0.98   \\
  ESC &   0.13 &   0.09 &  -0.03 &   0.04 &   0.17 &   1.00 &  &  -0.09 &   0.66 &   1.18 &  -2.17 &   2.50 &   0.08 &   0.09 &   1.11   \\
  MPO &   0.08 &   0.15 &  -0.03 &   0.11 &   0.16 &   1.12 &  &  -0.04 &   0.37 &   0.03 &  -1.50 &   0.01 &   0.09 &   0.12 &   1.11   \\
  MP1 &   0.08 &   0.20 &  -0.16 &   0.06 &   0.13 &   1.01 &  &  -0.17 &   0.41 &  -0.34 &  -1.11 &   0.22 &   0.05 &   0.11 &   1.04   \\
  RW &   0.11 &   0.11 &  -0.11 &   0.15 &   0.13 &   1.04 &  &  -0.16 &   0.48 &   1.30 &  -1.92 &   2.20 &   0.09 &   0.07 &   1.09   \\
 TAR &   0.14 &   0.14 &  -0.11 &   0.10 &   0.11 &   1.01 &  &  -0.11 &   0.58 &   0.23 &  -0.87 &  -1.30 &   0.08 &   0.08 &   1.11   \\
 PCO &   0.14 &   0.06 &   0.12 &   0.04 &   0.17 &   1.37 &  &   0.08 &   0.26 &   1.41 &  -1.35 &  -1.34 &   0.03 &   0.16 &   1.65   \\
 PFO &   0.22 &   0.06 &   0.19 &   0.02 &   0.15 &   1.26 &  &   0.17 &   0.08 &   0.12 &  -1.84 &   0.77 &   0.02 &   0.13 &   1.32   \\
 PCR &   0.13 &   0.06 &  -0.74 &   0.16 &   0.37 &   1.27 &  &  -0.59 &   3.48 &  -3.18 &   2.54 &  -4.77 &   0.26 &   0.28 &   1.16   \\
 PFR &   0.22 &   0.06 &  -0.31 &   0.28 &   0.13 &   0.59 &  &  -0.26 &   1.77 &  -0.59 &   0.19 &  -1.51 &   0.15 &   0.08 &   0.73   \\
 MVM &   0.13 &   0.11 &   0.06 &   0.08 &   0.06 &   0.63 &  &   0.07 &   0.40 &  -0.65 &  -0.19 &  -0.87 &   0.06 &   0.06 &   0.75   \\
 AS1 &   0.11 &   0.31 &   0.09 &   0.23 &   0.18 &   0.97 &  &   0.09 &   1.24 &  -0.18 &  -0.10 &   0.32 &   0.20 &   0.17 &   1.11   \\
 AS2 &   0.11 &   0.31 &   0.18 &   0.24 &   0.18 &   0.86 &  &   0.16 &   1.17 &  -0.18 &  -0.01 &   0.52 &   0.20 &   0.17 &   0.98   \\
 AvL &   0.11 &   0.10 &   0.16 &   0.15 &   0.13 &   1.01 &  &   0.13 &   0.46 &   0.18 &  -2.31 &   1.93 &   0.03 &   0.08 &   1.05   \\
 CLE &   0.13 &   0.11 &  -0.11 &   0.06 &   0.17 &   0.97 &  &  -0.18 &   0.58 &   0.70 &  -2.11 &   1.82 &   0.02 &   0.10 &   0.98   \\
 CLN &   0.08 &   0.15 &  -0.24 &   0.08 &   0.14 &   1.06 &  &  -0.24 &   0.29 &  -0.49 &  -1.55 &  -0.59 &   0.02 &   0.09 &   1.02   \\
 SG1 &   0.18 &   0.08 &   0.07 &   0.03 &   0.20 &   0.92 &  &  -0.01 &   0.97 &   1.19 &  -1.66 &   0.50 &   0.02 &   0.12 &   1.01   \\
 SG2 &   0.10 &   0.20 &  -0.15 &   0.07 &   0.14 &   0.94 &  &  -0.20 &   0.58 &  -0.13 &  -1.18 &   1.16 &   0.02 &   0.11 &   1.03   \\
 SG3 &   0.25 &   0.07 &  -0.05 &   0.05 &   0.12 &   1.04 &  &  -0.08 &   0.39 &   0.33 &  -1.36 &   1.00 &   0.02 &   0.10 &   1.06   \\
 PCS &   0.13 &   0.06 &  -0.17 &   0.16 &   0.12 &   1.03 &  &  -0.21 &   0.32 &   0.73 &  -1.47 &   1.19 &   0.16 &   0.11 &   1.34   \\
 PFS &   0.22 &   0.06 &  -0.15 &   0.16 &   0.12 &   1.08 &  &  -0.19 &  -0.04 &   0.96 &  -1.16 &  -0.42 &   0.12 &   0.12 &   1.07   \\
\hline
    \end{tabular}
    \caption{Best fit parameters of a model describing dependance of cluster mass estimates on true cluster mass, contamination and 
    incompleteness, as given by eq.~(\ref{mu_model_g}). `Simplistic' columns show results for a restricted model with no dependance on 
    contamination and incompleteness ($\mu_{C1}=\mu_{I1}=\mu_{C2}=\mu_{I2}=0$), mean contamination ($\langle C\rangle$) and 
    mean incompleteness ($\langle I \rangle$). Parameter $\alpha_{0}$ is the slope of the $M_{\rm rec}-M_{\rm true}$ scaling relation, 
    parameters $\mu_{ij}$ are the coefficients of the linear ($j=1$) and quadratic ($j=2$) terms in contamination ($i=C$) or incompleteness 
    ($i=I$), while parameters $\sigma_{0}$ and $\sigma_{1}$ are respectively the intrinsic scatter and the Poisson-like scatter for $N_{\rm true}=100$ 
    cluster members (see eq.~[\ref{scatter}]). The model accounting for imperfect membership is strongly favored for every method with 
    $\Delta\rm{BIC}\ll-10$.}
    \label{tab:best_fit}
  \end{table*}

\end{document}